\documentclass[twocolumn,showpacs,superscriptaddress,preprintnumbers,prd]{revtex4}
\usepackage{graphicx,amsmath,amssymb,dcolumn,bm}
\begin{document}
%%%%%%%%%%%%%%%%%%%%%%%%%%%%%%%%%%%%%%%%%%%%%%%%%%%%%%%%%%%%%%%%%%%%%%%%%%%%%%%%
\preprint{KEK-TH-1080}
\title{\Large \bf Determination of polarized parton distribution functions \\
                  with recent data on polarization asymmetries}
\author{M. Hirai}
\email[E-mail: ]{mhirai@post.kek.jp}
\affiliation{Institute of Particle and Nuclear Studies \\
          High Energy Accelerator Research Organization (KEK) \\
          1-1, Ooho, Tsukuba, Ibaraki, 305-0801, Japan}
\author{S. Kumano}
\email[E-mail: ]{shunzo.kumano@kek.jp}
\affiliation{Institute of Particle and Nuclear Studies \\
          High Energy Accelerator Research Organization (KEK) \\
          1-1, Ooho, Tsukuba, Ibaraki, 305-0801, Japan}
\affiliation{Department of Particle and Nuclear Studies \\
           The Graduate University for Advanced Studies \\
           1-1, Ooho, Tsukuba, Ibaraki, 305-0801, Japan}     
\author{N. Saito}
\email[E-mail: ]{saito@nh.scphys.kyoto-u.ac.jp}
\affiliation{Department of Physics, Kyoto University, 
         Kyoto, 606-8502, Japan}
\collaboration{Asymmetry Analysis Collaboration}
% \homepage[URL: ]{http://spin.riken.bnl.gov/aac/}
% \date{\today}
\date{March 25, 2006}
%%%%%%%%%%%%%%%%%%%%%%%%%%%%%%%%%%%%%%%%%%%%%%%%%%%%%%%%%%%%%%%%%%%%%%%%%%%%%%%%
\begin{abstract}
Global analysis has been performed within the next-to-leading order
in Quantum Chromodynamics (QCD) to determine polarized parton distributions
with new experimental data in spin asymmetries. The new data set includes
JLab, {\sc Hermes}, and {\sc Compass} measurements on spin asymmetry $A_1$
for the neutron and deuteron in lepton scattering. Our new analysis also
utilizes the double-spin asymmetry for $\pi^0$ production in polarized
$pp$ collisions, $A_{LL}^{\pi^0}$, measured by the {\sc Phenix} collaboration.
Because of these new data, uncertainties of the polarized PDFs are
reduced. In particular, the JLab, {\sc Hermes}, and {\sc Compass}
measurements are valuable for determining $\Delta d_v(x)$ at large $x$
and $\Delta \bar q(x)$ at $x \sim 0.1$. The {\sc Phenix} $\pi^0$ data
significantly reduce the uncertainty of $\Delta g(x)$. Furthermore,
we discuss a possible constraint on $\Delta g(x)$ at large $x$ by
using the {\sc Hermes} data on $g_1^d$ in comparison with the {\sc Compass}
ones at $x\sim 0.05$. 
\end{abstract}

%%%%%%%%%%%%%%%%%%%%%%%%%%%%%%%%%%%%%%%%%%%%%%%%%%%%%%%%%%%%%%%%%%%%%%%%%%%%%%%%
\maketitle

%%%%%%%%%%%%%%%%%%%%%%%%%%%%%%%%%%%%%%%%%%%%%%%%%%%%%%%%%%%%%%%%%%%%%%%%%%%%%%%%
%%%%%%%%%%%%%%%%%%%%%%%%%%%%%%%%%%%%%%%%%%%%%%%%%%%%%%%%%%%%%%%%%%%%%%%%%%%%%%%%
\section{Introduction}

The nucleon spin structure has been investigated mainly by polarized
lepton-nucleon scattering experiments. After many years of theoretical
and experimental efforts, our knowledge on the spin structure of
the nucleon is much improved especially in the valence-quark sector.
However, we still miss a decent measurement of orbital angular momenta,
transversity distributions, and helicity distributions of sea-quarks
and gluons in the proton.
In particular, the polarized gluon distribution has large uncertainty.
For example. the AAC analysis indicated that the first moment of
$\Delta g(x)$ is $\Delta g =0.50 \pm 1.27$ at $Q^2$=1~GeV$^2$
\cite{aac00,aac03}. This represents the largest ambiguity in
understanding the longitudinal spin-structure of the proton;
$\Delta g$, the gluon spin contribution to the proton can be
either positive or negative.

Fortunately, various experiments started to produce data
which are sensitive to the gluon polarization. First, there were reports
on $\Delta g(x)/g(x)$ from {\sc Hermes} \cite{hermes-dg}, 
SMC (Spin Muon Collaboration) \cite{smc-dg}, and {\sc Compass}
\cite{compass-dg-lowq2} collaborations by using high-$p_T$
hadron production measurements in polarized lepton-nucleon scattering.
There are also preliminary results \cite{compass-dg-preli} from
the {\sc Compass} by using the data for charmed-meson and high-$p_T$-hadron
productions. Second, the RHIC (Relativistic Heavy Ion Collider) Spin
experiments started to produce data which could constrain the gluon
polarization. They are spin-asymmetry measurements in pion
\cite{phenix-pi-04,phenix-pi-06} and jet \cite{star-jet} productions
in polarized $pp$ reactions by {\sc Phenix} and {\sc Star} collaborations,
respectively. Because of these new data, we expect much improved
determination of the polarized gluon distribution.

There are also new measurements on the spin asymmetry $A_1$ in 
lepton-nucleon scattering by JLab (Thomas Jefferson National
Accelerator Facility) Hall A \cite{jlab-g1n}, {\sc Hermes} \cite{hermes-g1d},
and {\sc Compass} \cite{compass-g1d} collaborations in addition to
the data set used in our previous analysis \cite{aac03}. These data should
be useful for a better determination of quark and anti-quark distributions,
and they may also provide a constraint on the gluon polarization through
scaling violation. In addition, the {\sc Phenix} $\pi^0$ data can be
included in the analysis for a better determination of the polarized
gluon distribution. 

The major updates from the previous analysis \cite{aac03} is the addition of
these new data. Using these reaction data, we determine the polarized parton 
distribution functions (PDFs). We discuss analysis results by demonstrating
the importance of each experiment. In particular, new points are
the discussions on:
\begin{enumerate}
\item impact of the JLab, {\sc Hermes}, and {\sc Compass} data
      on the determination of $\Delta q(x)$ and $\Delta \bar q(x)$,
\item impact of {\sc Phenix} data on the determination of $\Delta g(x)$,
\item constraint on $\Delta g(x)$ from the differences between {\sc Hermes}
    and {\sc Compass} data on $g_1^d$ at $x\sim 0.05$.
\end{enumerate}

This paper is organized as follows. In Sec. \ref{analysis}, our analysis
method is described for determining the polarized PDFs. Analysis results
are explained in Sec. \ref{results}. The parametrization results are
compared with recent spin asymmetry data, the optimum polarized PDFs
are shown, and they are discussed in comparison with other PDFs.
Then, details of the polarized gluon distribution is discussed
in Sec. \ref{gluon}. The results are summarized in Sec. \ref{summary}.

%%%%%%%%%%%%%%%%%%%%%%%%%%%%%%%%%%%%%%%%%%%%%%%%%%%%%%%%%%%%%%%%%%%%%%%%%%%%%%%%
%%%%%%%%%%%%%%%%%%%%%%%%%%%%%%%%%%%%%%%%%%%%%%%%%%%%%%%%%%%%%%%%%%%%%%%%%%%%%%%%
\section{\label{analysis} Analysis method}

Spin-asymmetry data are analyzed to obtain the polarized PDFs.
In lepton-nucleon deep inelastic scattering (DIS), the spin asymmetry $A_1$
is given by unpolarized structure function $F_2$, longitudinal-transverse
structure function ratio $R$, and polarized structure function $g_1$
\cite{a1}:
\begin{equation}
        A_1(x, Q^2)=\frac{g_1(x, Q^2)}{F_2(x, Q^2)}\,
               2 \, x \, [1+R(x, Q^2)] \, ,
\label{eqn:a1}
\end{equation}
where $Q^2$ is given by the momentum transfer $q$ as $Q^2=-q^2$,
and $x$ is the Bjorken scaling variable defined by 
$x=Q^2/(2p\cdot q)$ with the nucleon momentum $p$.
The structure function $g_1$ is expressed \cite{g1}:
\begin{align}
g_1 (x, & Q^2) = \frac{1}{2}\sum\limits_{i=1}^{n_f} e_{i}^2
   \bigg\{ \Delta C_q(x,\alpha_s) \otimes [ \Delta q_{i} (x,Q^2)
\nonumber \\
   & + \Delta \bar{q}_{i} (x,Q^2) ]
    + \Delta C_g(x,\alpha_s) \otimes \Delta g (x,Q^2) \bigg\},
\label{eqn:g1}
\end{align}
where $\Delta C_q$ and $\Delta C_g$ are polarized coefficient functions,
and $\Delta q_{i}$, $\Delta \bar{q}_{i}$, and $\Delta g$ are polarized
quark, antiquark, and gluon distribution functions, respectively.
The symbol $\otimes$ denotes the convolution integral:
\begin{equation}
f (x) \otimes g (x) = \int^{1}_{x} \frac{dz}{z}
            f(z) g\left(\frac{x}{z}\right) .
\label{eqn:conv}
\end{equation}
In the similar way, the structure function $F_2$ is expressed in terms of
unpolarized PDFs and coefficient functions.
The structure functions are calculated in the next-to-leading-order (NLO)
of the running coupling constant $\alpha_s$, and the modified minimal
subtraction ($\overline {\rm MS}$) scheme is used.

The longitudinal double spin asymmetry for pion production 
in polarized $pp$ reaction is given by \cite{phenix-pi-04,phenix-pi-06}
\begin{equation}
A_{LL}^{\pi^0}=\frac{[d\sigma_{++}-d\sigma_{+-}]/dp_T}
                    {[d\sigma_{++}+d\sigma_{+-}]/dp_T}
              =\frac{d\Delta\sigma/dp_T}{d\sigma/dp_T} ,
\end{equation}
where $p_T$ is the transverse momentum of the pion, and $\sigma_{h_1 h_2}$
is the cross section with the proton helicities $h_1$ and $h_2$. 
The cross sections $\Delta\sigma$ and $\sigma$ are defined by
$\Delta\sigma=(\sigma_{++}-\sigma_{+-})/2$ and 
$\sigma=(\sigma_{++}+\sigma_{+-})/2$. 
The polarized cross sections are expressed in terms of the polarized PDFs
$\Delta f$ and fragmentation functions $D_c^{\pi^0}$
\cite{a1-pi-theo}:
\begin{align}
& \! \! \! \frac{{d \Delta \sigma }}{{dp_T }} =
\sum\limits_{a,b,c} {\int_{\eta _{\min } }^{\eta _{\max } } {d\eta }
\int_{x_a^{\min } }^1 {dx_a } } \Delta f_a (x_a )
\nonumber \\
& \times
\int_{x_b^{\min } }^1 {dx_b } \Delta f_b (x_b )
\frac{{\partial (\hat t,z_c )}}{{\partial (p_T ,\eta )}}
\frac{{d\Delta \hat \sigma_{a+b\rightarrow c+X}}}{{d\hat t }} D_c^\pi  (z_c ) ,
\label{pol-cross}
\end{align}
where $\partial (\hat t,z_c )/\partial (p_T ,\eta )$ is the Jacobian.
The cross section $d\Delta \hat \sigma_{a+b\rightarrow c+X} /d \hat t$ is 
for the parton-level subprocess $a+b\rightarrow c+X$, and its expression is,
for example, found in Ref. \cite{subprocess}.
The variables $x_a$ and $x_b$ are momentum fractions
for the partons $a$ and $b$, respectively, and $z_c$
is the momentum fraction given by $z_c=p_\pi / p_c$
with the pion momentum $p_\pi$ and the parton momentum $p_c$.
The pseudorapidity $\eta$ and $\hat t$ are defined by 
\begin{equation}
\eta = - \ln \left[ {\tan \left( {\frac{{\theta_\pi}}{2}} \right)} \right] ,
\ \ \hat t = (p_a - p_c)^2 ,
\end{equation}
where $\theta_\pi$ is the angle from the beam direction in the laboratory
frame. The bounds of the integral over $\eta$, $\eta_{max}$ and $\eta_{min}$,
are given by the experimental condition. 
Using Madelstam variables $s=(p_A+p_B)^2$,
$t=(p_A-p_\pi)^2$, and $u~=(p_B-p_\pi)^2$, we define variables
$x_1$ and $x_2$ \cite{a1-pi-kine}:
\begin{equation}
x_1  \equiv -\frac{u}{s} = \frac{x_T}{2}e^{ + \eta } , \ \ 
x_2  \equiv -\frac{t}{s} = \frac{x_T}{2}e^{ - \eta } ,
\end{equation}
where $x_T$ is given by $x_T  = 2p_T / \sqrt s$. The lower bounds of
the integrals and the variable $z_c$ are then expressed:
\begin{equation}
x_a^{\min }  = \frac{x_1}{1 - x_2}, \ \ 
x_b^{\min }  = \frac{x_a x_2}{x_a  - x_1}, \ \ 
z_c  = \frac{x_1}{x_a} + \frac{x_2}{x_b}.
\end{equation}

The polarized cross section in  Eq. (\ref{pol-cross}) is calculated
by these expressions together with polarized PDFs and fragmentation
functions. The unpolarized cross section $d\sigma/dp_T$ is calculated
in the similar way. In calculating the NLO cross sections, $K$-factors
are simply multiplied in order to obtain them from the leading-order (LO)
ones as was done in unpolarized PDF analyses \cite{CTEQ-K}. 
We employed $K$=1.0 and 1.6 for the polarized and unpolarized cross sections,
respectively, basing on more detailed analyses in Ref. \cite{a1-pi-theo}.
This approximation inevitably introduces systematic uncertainties from
the $p_T$ dependence of the $K$ factor and the choice of the PDF set.
Such an error is estimated to be $\sim$20\% in the cross sections,
which is comparable magnitude to a typical scale uncertainty
in this energy region. A possible criticism to $K$-factor approach
in PDF analysis might be a change in corresponding $x$-region due to
higher-order corrections, which should be the origin of the $K$-factor.
Since the cross section drops rapidly with increasing $p_T$, which is
a good measure of relevant $x$ region especially for the {\sc Phenix}
acceptance covering only central rapidity, changes in the cross
section would correspond to small change in $p_T$ even if the change is
fully attributed to a shift in $p_T$. For example, 50\% changes in the
cross section can be obtained by 10\% change in $p_T$. Therefore, we
think the uncertainty involved in the $x$-determination due to $K$-factor
approach is smaller compared to other uncertainties. Clearly more precise
procedure has to be applied as the precision of experimental data improves.

The polarized PDFs are expressed by a number of parameters and unpolarized
PDFs at a fixed $Q^2$ ($\equiv Q_0^2$):
\begin{equation}
   \Delta f(x,Q_0^2) 
   = [\delta x^{\nu}-\kappa (x^{\nu}-x^{\mu})] f(x,Q_0^2) \, ,
\label{eqn:df}
\end{equation}
where $\delta$, $\kappa$, $\nu$, and $\mu$ are parameters to be determined
by a $\chi^2$ analysis, and $f(x)$ is the corresponding unpolarized PDF.
Assuming flavor-symmetric polarized antiquark distributions
$\Delta \bar u = \Delta \bar d = \Delta \bar s \equiv \Delta \bar q$
at $Q_0^2$, we determine four polarized distributions, $\Delta u_v$,
$\Delta d_v$, $\Delta \bar q$, and $\Delta g$, which are expressed by
the parameters in Eq. (\ref{eqn:df}). As usual in this kind of analysis,
the first moments of $\Delta u_v$ and $\Delta d_v$ are fixed
by semileptonic decay data of octet baryons:
$\int dx \Delta u_v=0.926$, $\int dx \Delta d_v =-0.341$ \cite{aac00,aac03}. 

The positivity condition $|\Delta f(x)| \le f(x)$ is imposed in the initial
PDFs. It should be satisfied in the LO, whereas it does not have to be
strictly satisfied in the NLO due to NLO corrections. 
The details of the NLO corrections on the positivity condition are
discussed in Ref. \cite{afr98}. We have imposed the condition only
for a practical reason to avoid unphysical solutions particularly
at large $x$. When the precision of experimental data is improved,
we should be able to release this condition. A recent unpolarized
gluon distribution \cite{CTEQ-K,otherpdfs} becomes negative
at small $x$ ($<0.01$) and $Q^2$ ($\sim$1 GeV$^2$), a special
attention should be paid to the condition. In such a region, few
polarized data are used in our analysis at this stage, so that
the positivity condition does not affect the analysis significantly.

The parameters of the polarized PDFs are determined so as to minimize
the total $\chi^2$:
\begin{equation}
\chi^2=\sum_i \frac{[ A_i^{\rm data}(x,Q^2)
                     -A_i^{\rm calc}(x,Q^2) ]^2}
                {[\Delta A_i^{\rm data}(x,Q^2) ]^2} ,
\label{eqn:chi2}
\end{equation}
where $A_i^{\rm data}$ indicates the spin-asymmetry data for
$A_1$ and $A_{LL}^{\pi^0}$; $A_i^{\rm calc}$ is the theoretically
calculated asymmetry at the same $Q^2$ point. The polarized and unpolarized
PDFs at $Q_0^2$ are evolved to the experimental $Q^2$ point by the standard
DGLAP (Dokshitzer-Gribov-Lipatov-Altarelli-Parisi) evolution equations
\cite{dglap-evol}. The experimental error $\Delta A_i^{\rm data}$ is given
by systematic and statistical errors:
$(\Delta A_i^{\rm data})^2
     =(\Delta A_i^{\rm stat})^2+(\Delta A_i^{\rm syst})^2$.
The total $\chi^2$ is minimized by the CERN program library {\tt MINUIT}
\cite{minuit}. 

The uncertainties of the polarized PDFs are estimated by the Hessian method
\cite{hessian}.
We express the parameters as $a_i$ ($i$=1, 2, ..., $N$), where $N$ is
the number of the parameters, and the minimum point of $\chi^2$ is denoted
$\hat a$. Then, the uncertainty of a distribution $F(x)$ is calculated by
\begin{equation}
        [\delta F(x)]^2=\Delta \chi^2 \sum_{i,j}
          \left( \frac{\partial F(x,\hat{a})}{\partial a_i}  \right)
          H_{ij}^{-1}
          \left( \frac{\partial F(x,\hat{a})}{\partial a_j}  \right) \ ,
        \label{eq:erroe-M}
\end{equation}
where $H_{ij}$ is the Hessian. The value of $\Delta \chi^2$ is given
by $\Delta \chi^2=12.65$ so that the uncertainty indicates
the one-$\sigma$-error range for eleven free parameters \cite{aac03}.
The details of this $\Delta \chi^2$ choice are discussed in a number
of publications \cite{del-chi2}.

%%%%%%%%%%%%%%%%%%%%%%%%%%%%%%%%%%%%%%%%%%%%%%%%%%%%%%%%%%%%%%%%%%%%%%%%%%%%%%%%
%%%%%%%%%%%%%%%%%%%%%%%%%%%%%%%%%%%%%%%%%%%%%%%%%%%%%%%%%%%%%%%%%%%%%%%%%%%%%%%%
\section{\label{results} Results}

We determine the parameters by minimizing the total $\chi^2$
in Eq. (\ref{eqn:chi2}). In order to find an impact 
of the RHIC-$A_{LL}^{\pi^0}$ data on the polarized PDF determination,
we tried two types of analyses:
\begin{itemize}
\item
    Type 1: with the DIS-$A_1$ and RHIC-$A_{LL}^{\pi^0}$ data,
\item
    Type 2: with only the DIS-$A_1$ data.
\end{itemize}
For the theoretical $\pi^0$ production calculations, the kinematical
conditions, $\sqrt{s}$=200 GeV and $|\eta| \le$0.35, are used.
The initial $Q^2$ point is taken as $Q_0^2 = 1$~GeV$^2$.
We have tried another fit with $Q_0^2 = 0.8$~GeV$^2$, but
there was no difference in the results. 
The number of flavor is three. The longitudinal-transverse ratio $R$
in Eq. (\ref{eqn:a1}) is taken from the SLAC parametrization \cite{SLAC-R}.
The GRV98 parametrization \cite{GRV98} is used for the unpolarized PDFs.
We also tested other distributions; however, the obtained polarized PDFs
did not change conspicuously. It is because the differences in the PDFs
could be absorbed into the parametrized function in Eq. (\ref{eqn:df})
by minor adjustments of the parameters.

%%%%%%%%%%%%%%%%%%%%%%% This table should be in Sec.3 %%%%%%%%%%%%%%%%%%%%%%%
%%%%%%%%%%%%%%%%%%%%%%%%%%%%%%%% table  %%%%%%%%%%%%%%%%%%%%%%%%%%%%%%%%%%%%%%%%
\begin{table*}
\caption{\label{T:NLO_pi0p}
Obtained parameters by the $\chi^2$ analyses.}
\begin{ruledtabular}
\begin{tabular}{ccccc} 
distribution \  & $\delta$             & $\nu$
                & $\kappa$             & $\mu$  \\
\hline
Type 1: DIS and $\pi^0$ data
                &                      &                   \\
$\Delta u_v$    &    0.959 $\pm$ 0.099 & 0.0 (fixed)     
                &    0.588             & 1.048 $\pm$ 0.266 \\
$\Delta d_v$    & $-$0.773 $\pm$ 0.210 & 0.0 (fixed)     
                & $-$0.478             & 1.243 $\pm$ 0.561 \\
$\Delta \bar q$ &    0.780 $\pm$ 0.904 & 1.014 $\pm$ 0.180 
                & $-$75.9  $\pm$ 11.2  & 1.0 (fixed)     \\
$\Delta g$      & $-$1.00  $\pm$ 4.39  & 2.74  $\pm$ 1.30 
                &    252   $\pm$ 139   & 2.70  $\pm$ 2.57 \\
\hline
Type 2: DIS data only
                &                      &                   \\
$\Delta u_v$    &    0.958 $\pm$ 0.101 & 0.0 (fixed)     
                &    0.585             & 1.056 $\pm$ 0.272 \\
$\Delta d_v$    & $-$0.768 $\pm$ 0.204 & 0.0 (fixed)     
                & $-$0.474             & 1.218 $\pm$ 0.552 \\
$\Delta \bar q$ &    0.955 $\pm$ 0.951 & 1.014 $\pm$ 0.185 
                & $-$82.7  $\pm$ 13.2  & 1.0 (fixed)       \\
$\Delta g$      & $-$1.00  $\pm$ 3.28  & 2.30  $\pm$ 0.883 
                &    253   $\pm$ 153   & 2.27  $\pm$ 1.75 
\end{tabular}
\end{ruledtabular}
\end{table*}
%%%%%%%%%%%%%%%%%%%%%%%%%%%%%%%% table  %%%%%%%%%%%%%%%%%%%%%%%%%%%%%%%%%%%%%%%%

The obtained parameters are tabulated in Table \ref{T:NLO_pi0p}.
\footnote{One should note that the parameters are not
          determined within one percent accuracy.}
Their values are similar to those of the previous version \cite{aac03}.
The small-$x$ behavior of the polarized antiquark distributions cannot
be determined by the current experimental data, so that the parameter
$\mu_{\bar q}$, which determined the small-$x$ behavior, is fixed
at $\mu_{\bar q}=1$ as in Ref. \cite{aac03}. Because the first
moments of $\Delta u_v$ and $\Delta d_v$ are fixed, there are no errors
in the parameters $\kappa_{u_v}$ and $\kappa_{d_v}$.

%%%%%%%%%%%%%%%%%%%%%%%%%%%%%%%% table  %%%%%%%%%%%%%%%%%%%%%%%%%%%%%%%%%%%%%%%%
\begin{table*}[t]
\caption{\label{T:chi2}
Numbers of the $A_1$ data and $\chi^2$ values are listed for the two
types of analyses. The notations $p$, $n$, and $d$ indicate proton,
neutron, and deuteron, respectively.
}

\begin{ruledtabular}
\begin{tabular}{cccccc} 
data set       &  $x$ range  & $Q^2$ range & No. of data & 
                  \multicolumn{2}{c}{$\chi^2$} \\
               &             &  (GeV$^2$)  &             & 
                  Type 1              & Type 2       \\
               &             &             &             & 
                  DIS-$A_1$+$A_{LL}^{\pi^0}$  & DIS-$A_1$ only \\
\hline
EMC ($p$)           & 0.015$-$0.466 & 3.50$-$29.5  
                          & 10  &   4.84  &   4.81   \\
SMC ($p$)           & 0.004$-$0.484 & 1.14$-$72.10 
                          & 59  &  55.72  &  55.21   \\
E130 ($p$)          & 0.19$-$0.64   & 5.32$-$9.91   
                          & 8   &   4.74  &   4.72   \\
E143 ($p$)          & 0.027$-$0.749 & 1.17$-$9.52  
                          & 81  &  61.56  &  60.78   \\
E155 ($p$)          & 0.015$-$0.750 & 1.22$-$34.72 
                          & 24  &  32.41  &  32.88   \\
{\sc Hermes} ($p$)  & 0.033$-$0.447 & 1.22$-$9.18 
                          &  9  &   3.32  &   3.31   \\
\hline                                            
SMC ($d$)           & 0.0042$-$0.483 & 1.14$-$71.76
                          & 65  &  57.68  &  57.16   \\
E143 ($d$)          & 0.027$-$0.749  & 1.17$-$9.52
                          & 81  &  88.33  &  89.13   \\
E155 ($d$)          & 0.015$-$0.750  & 1.22$-$34.79
                          & 24  &  18.22  &  18.26   \\
{\sc Hermes} ($d$)  & 0.033$-$0.446  & 1.22$-$9.16 
                          &  9  &  12.62  &  11.88   \\
{\sc Compass} ($d$) & 0.051$-$0.474  & 1.18$-$47.5
                          & 12  &   9.18  &   8.99   \\
\hline                                            
E142 ($n$)          & 0.035$-$0.466  & 1.1$-$5.5  
                          & 8   &   2.90  &   2.76   \\
E154 ($n$)          & 0.017$-$0.564  & 1.21$-$15.0
                          & 11  &   3.13  &   3.36   \\
{\sc Hermes} ($n$)  & 0.033$-$0.464  & 1.22$-$5.25
                          & 9   &   2.14  &   2.18   \\
J-Lab  ($n$)        & 0.33$-$0.60    & 2.71$-$4.83
                          & 3   &   2.50  &   2.63   \\
\hline
DIS total              & & & 413 & 359.29  & 358.05   \\
{\sc Phenix} ($\pi^0$) & & & 8   &  11.18  &   $-$    \\
\hline                                    
total $\chi^2$    & & & 421 & 370.47  & 358.05   \\
($\chi^2$/d.o.f.) & & &     & (0.904) & (0.891)
\end{tabular}
\end{ruledtabular}
\end{table*}
%%%%%%%%%%%%%%%%%%%%%%%%%%%%%%%% table  %%%%%%%%%%%%%%%%%%%%%%%%%%%%%%%%%%%%%%%%

The $\chi^2$ values for the used data set are listed in Table \ref{T:chi2}.
The data set includes the measurements from
EMC (European Muon Collaboration) \cite{emc}, SMC \cite{smc},
SLAC-E130, E143, E154, E155 \cite{slac},
{\sc Hermes} \cite{hermes97,hermes-g1d}, JLab Hall-A, \cite{jlab-g1n},
{\sc Compass} \cite{compass-g1d}, and {\sc Phenix} \cite{phenix-pi-06}
collaborations. The additional data, which were not used in the AAC03 analysis,
are the ones from JLab \cite{jlab-g1n}, {\sc Hermes} \cite{hermes-g1d},
{\sc Compass} \cite{compass-g1d}, and {\sc Phenix} \cite{phenix-pi-06}.
In two different analyses, similar $\chi^2/$d.o.f. values are obtained.
As we will see later, two fit results show almost the same $\Delta u_v(x)$,
and slight changes in the central curves of $\Delta d_v(x)$,
$\Delta \bar{q}(x)$, and $\Delta g(x)$. The major difference is in
the uncertainty bands for antiquark and gluon distributions.

%%%%%%%%%%%%%%%%%%%%%%%%%%%%%%%%%%%%%%%%%%%%%%%%%%%%%%%%%%%%%%%%%%%%%%%%%%%%%%%%
\subsection{\label{asym} Comparison with spin asymmetry measurements}

We compare the resulted parametrization with newly added experimental
data of DIS and $\pi^0$ production in $pp$ collisions. Since the
agreement with other data is found to be maintained, we do not show here.

\begin{figure}[t]
        \includegraphics*[width=60mm]{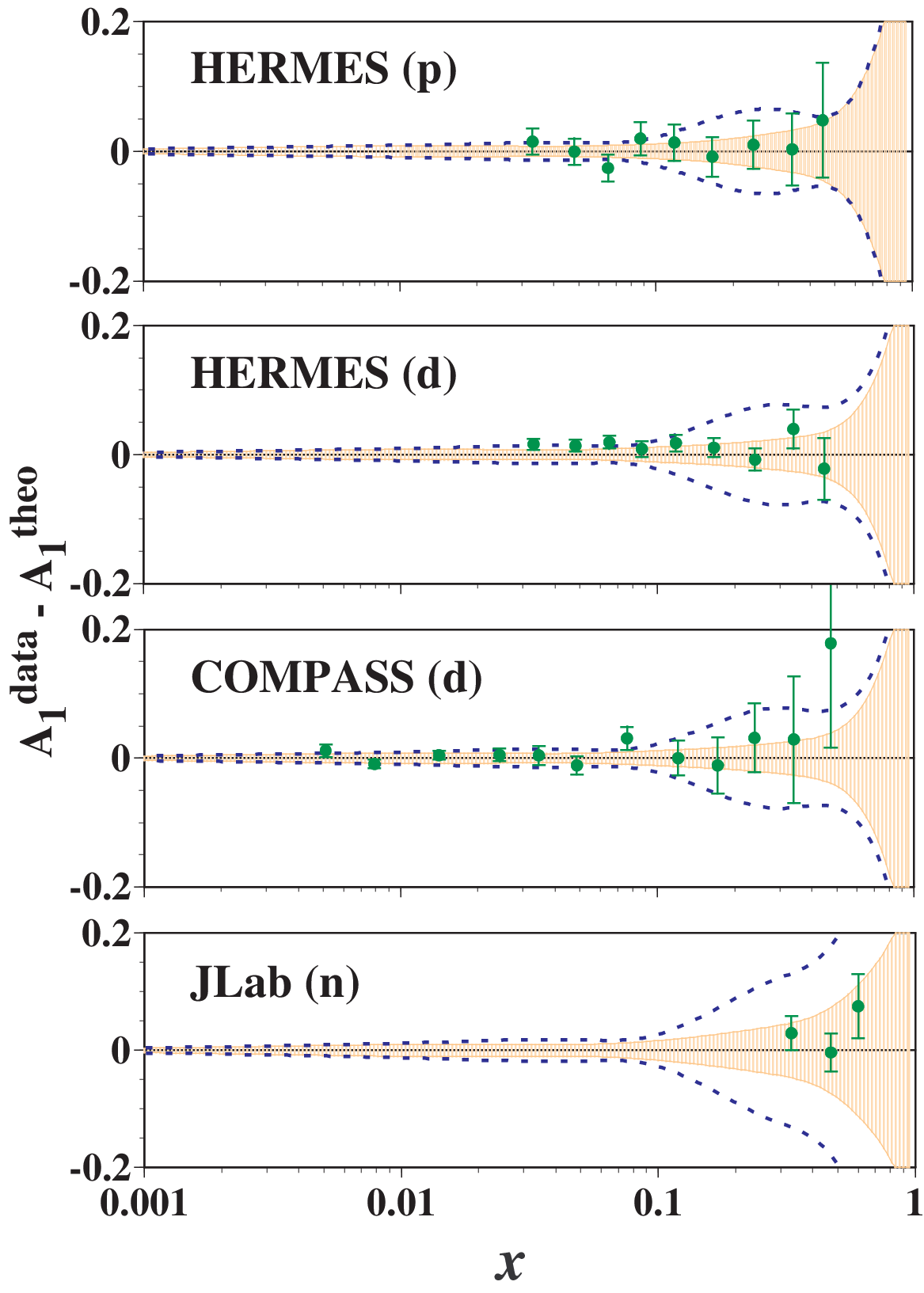}
\caption{\label{fig:a1diff} (Color online)
Differences between the data and parametrization results.
The dotted curves indicate uncertainty bands of the AAC03
analysis and the shaded areas are the bands of the current analysis.
}
\end{figure}

First, the parametrization results of the type-1 are compared with
inclusive DIS data by JLab, {\sc Hermes}, and {\sc Compass} on the spin
asymmetry $A_1$ in Fig. \ref{fig:a1diff}, where actual differences
between the theoretical asymmetries and experimental data
($A_1^{\rm data}-A_1^{\rm theo}$) are shown. The theoretical asymmetry
$A_1^{\rm theo}$ is calculated at the experimental $Q^2$ point of
$A_1^{\rm data}$. The AAC03 uncertainty bands are shown by the solid curves
and the current AAC06 type-1 uncertainty ranges are by the shaded areas.
The uncertainties are calculated at $Q^2$=5 GeV$^2$.
It is obvious from this figure that the {\sc Hermes} and {\sc Compass}
data on $A_1^p$ and $A_1^d$ significantly reduced (about 50\%) the
uncertainties of the asymmetries in the region, $x\sim 0.2$. On the other
hand, the JLab data play an important role in reducing the uncertainty
for the neutron in the region, $x\gtrsim 0.2$. These data are valuable
for a better determination of the polarized quark and antiquark
distributions as shown in Sec. \ref{pdfs}.

%%%%%%%%%%%%%%%%%%%%%%%%%%%%%%%% figure %%%%%%%%%%%%%%%%%%%%%%%%%%%%%%%%%%%%%%%%
%%%%%%%%%%%%%%%%%%%%%%%%%%%%%%%% figure %%%%%%%%%%%%%%%%%%%%%%%%%%%%%%%%%%%%%%%%

%%%%%%%%%%%%%%%%%%%%%%%%%%%%%%%% figure %%%%%%%%%%%%%%%%%%%%%%%%%%%%%%%%%%%%%%%%
\begin{figure}[t]
        \includegraphics*[width=70mm]{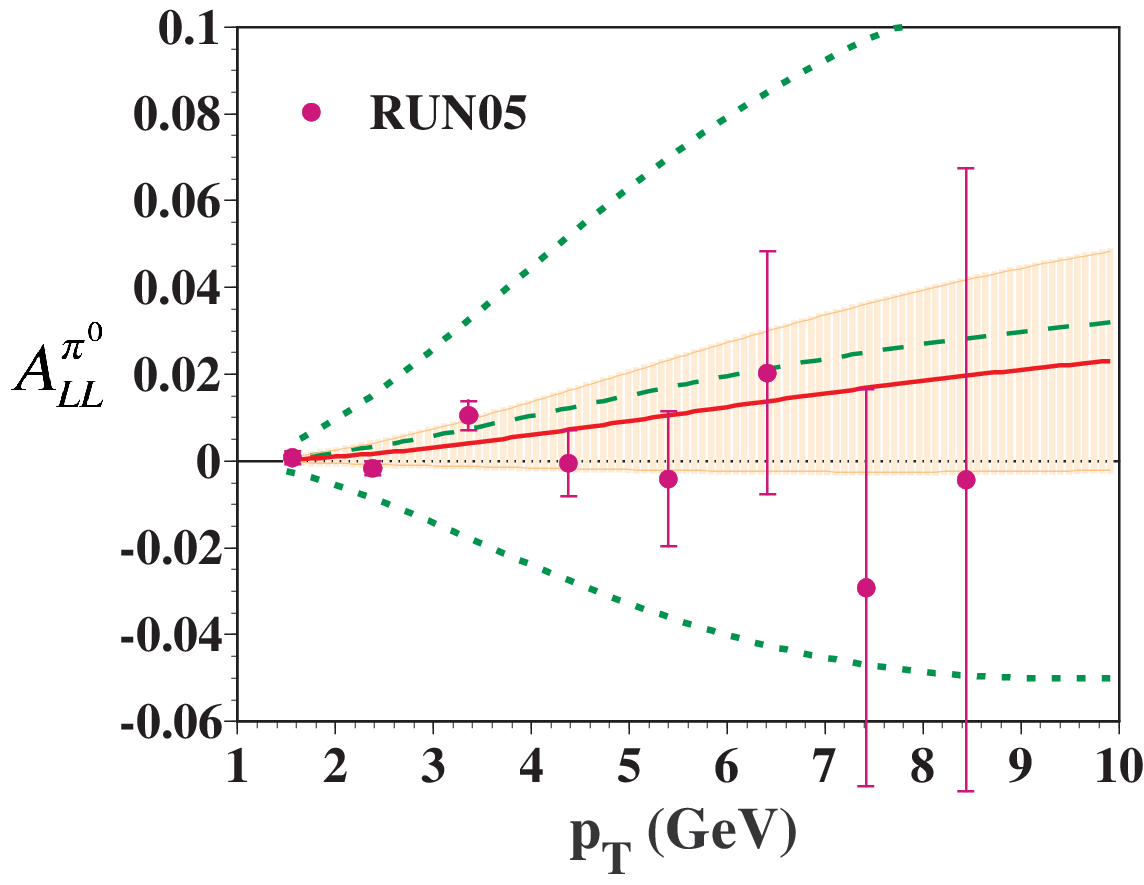}
\caption{\label{fig:asym_pi0} (Color online)
Comparison with {\sc Phenix} $\pi^0$-production asymmetry data $A_{LL}^{\pi^0}$.
The solid and dashed curves indicates type 1 and 2 analysis results,
respectively. The uncertainty range for the type 1 is shown by the shaded
band, and the one for the type 2 is by the dotted curves.
}
\end{figure}
%%%%%%%%%%%%%%%%%%%%%%%%%%%%%%%% figure %%%%%%%%%%%%%%%%%%%%%%%%%%%%%%%%%%%%%%%%

Second, the analysis results are compared with the {\sc Phenix} data
on the pion-production asymmetry $A_{LL}^{\pi^0}$ \cite{phenix-pi-06}
in Fig. \ref{fig:asym_pi0}. The theoretical curves and their uncertainties
are shown for the type 1 and 2 analyses.
The hard scale is taken as $Q^2$ = $p_T^2$ in calculating the PDF and
their uncertainties, and the KKP (Kniel, Kramer, and P{\"o}tter) fragmentation
functions \cite{kkp} are used. It is known from more detailed analysis that
the choice does not change the asymmetries \cite{RBRC-WS}.
It is clear from Fig. \ref{fig:asym_pi0} that the {\sc Phenix} data
should significantly improve the uncertainties in the PDFs, mainly
the polarized gluon distribution as we will see later.
However, the resulted PDFs are still consistent with
$A_{LL}^{\pi^0} = 0$ in the entire $p_T$ region, and it suggests a need
of more precise data. Nevertheless, significantly improved constraints
on the gluon distribution will be discussed basing on this fit results
in Sec. \ref{pdfs}.

%%%%%%%%%%%%%%%%%%%%%%%%%%%%%%%% figure %%%%%%%%%%%%%%%%%%%%%%%%%%%%%%%%%%%%%%%%
\begin{figure}[t]
        \includegraphics*[width=70mm]{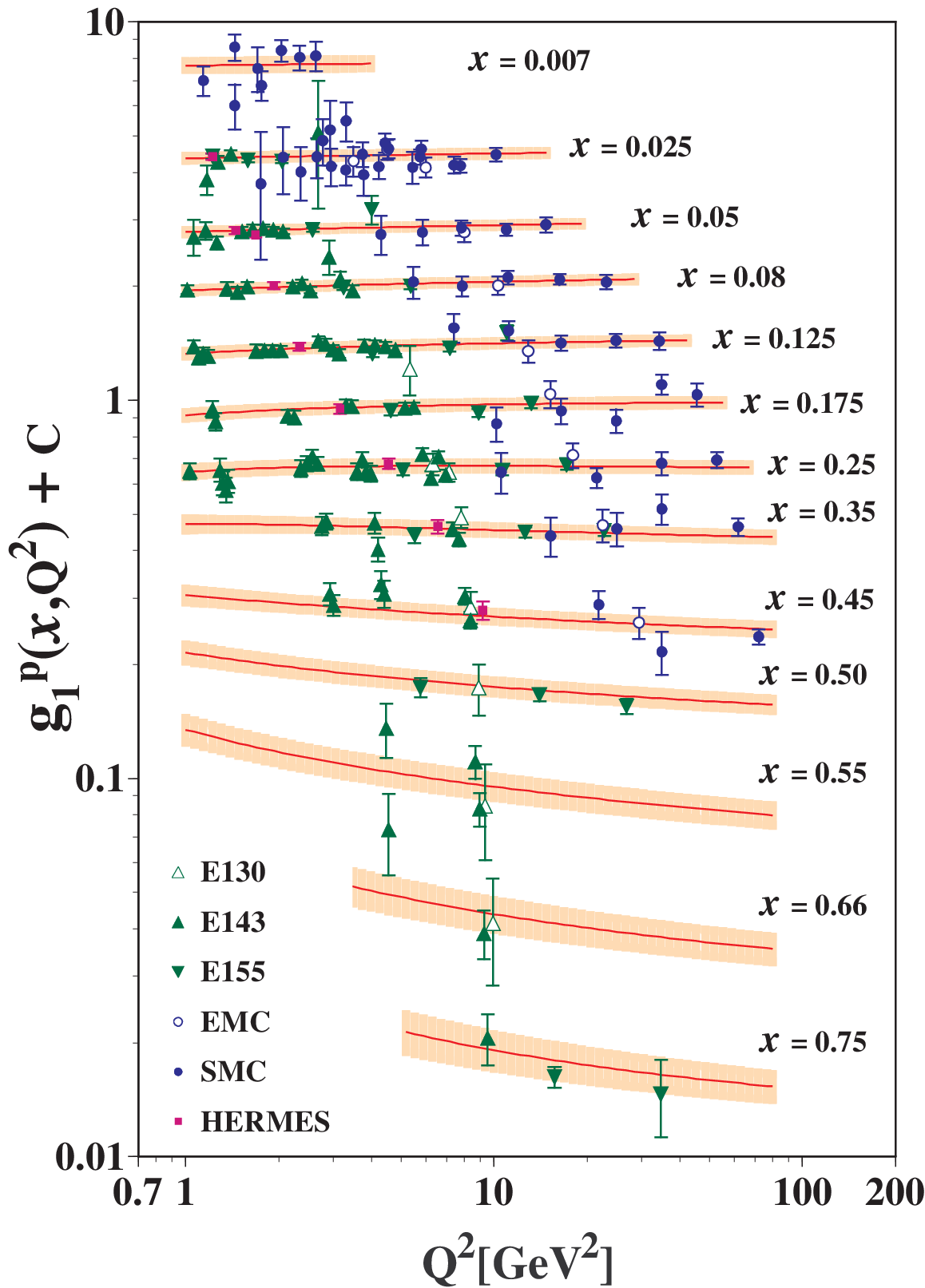}
\caption{\label{fig:g1} (Color online)
Current type-1 analysis results are compared with $g_1^p$ data.}
\end{figure}
%%%%%%%%%%%%%%%%%%%%%%%%%%%%%%%% figure %%%%%%%%%%%%%%%%%%%%%%%%%%%%%%%%%%%%%%%%

In order to illustrate the current precision of the structure function $g_1$
for the proton, the experimental data are compared with the fit results
including uncertainty bands in Fig. \ref{fig:g1}. Here, only the type-1 results
are shown. Experimental measurements are usually listed by the spin asymmetry
$A_1$, so that the $g_1^p$ data are calculated by using the unpolarized PDFs
of the GRV98 \cite{GRV98} and the SLAC parametrization for $R$ with
Eq. (\ref{eqn:a1}). We notice that the uncertainty bands are still wide
at large $x$ ($>$0.5). The smaller $x$ region ($x <0.007$) still remains
unmeasured. Although the kinematical coverage of the experimental data are being
extended, precision data are missing especially in the large-$x$ and extremely
small-$x$ regions, especially when we compared with the HERA data \cite{hera}.
Wider kinematical coverage is essential in extraction of $\Delta g$ through
a scaling violation. Such measurement would be feasible at the proposed
polarized $ep$-colliders, e-LIC \cite{elic} or eRHIC \cite{erhic}.

%%%%%%%%%%%%%%%%%%%%%%%%%%%%%%%%%%%%%%%%%%%%%%%%%%%%%%%%%%%%%%%%%%%%%%%%%%%%%%%%
\subsection{\label{pdfs} Polarized parton distribution functions}

\begin{figure}[t]
        \includegraphics*[width=40mm]{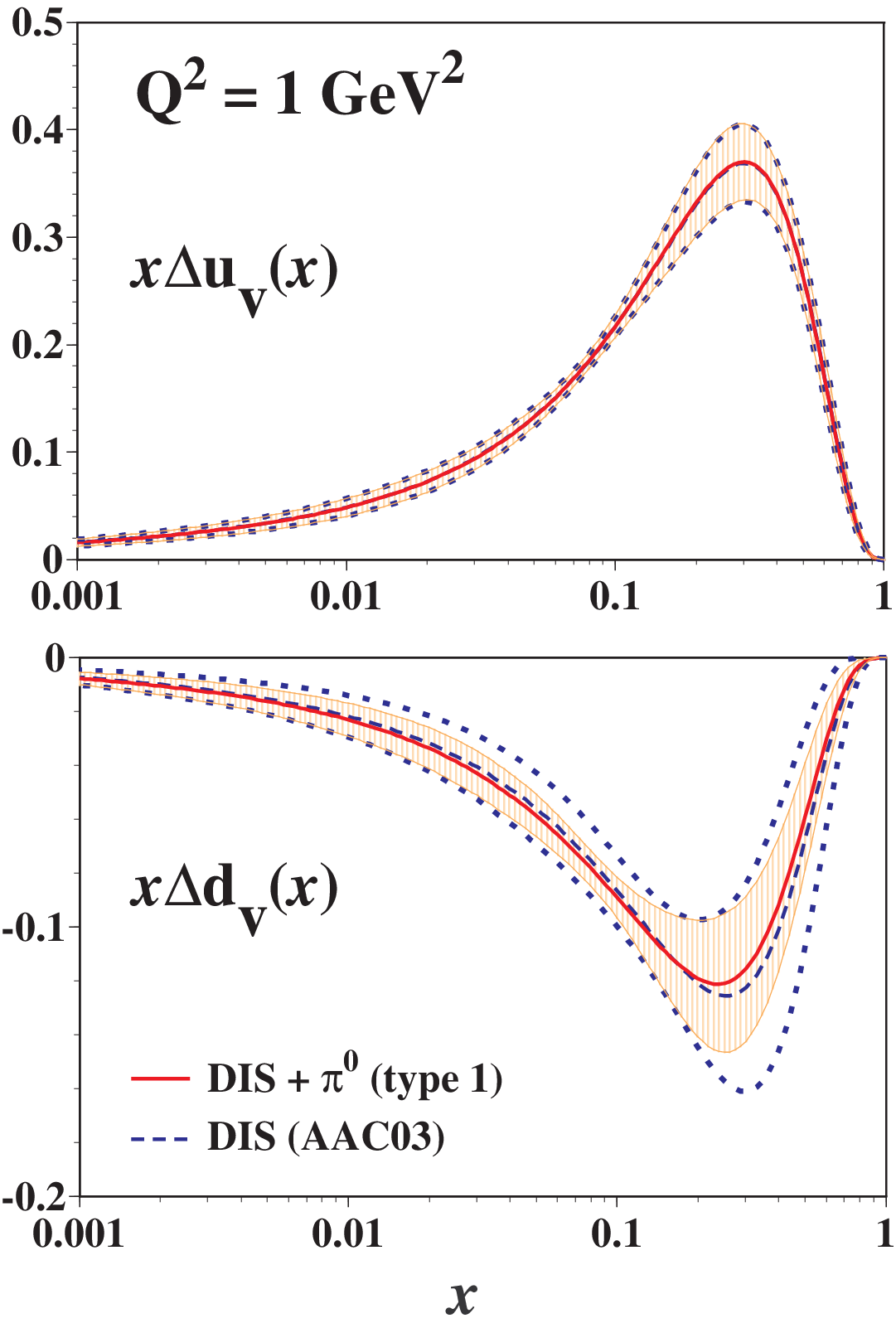} \hspace{1mm}
        \includegraphics*[width=41mm]{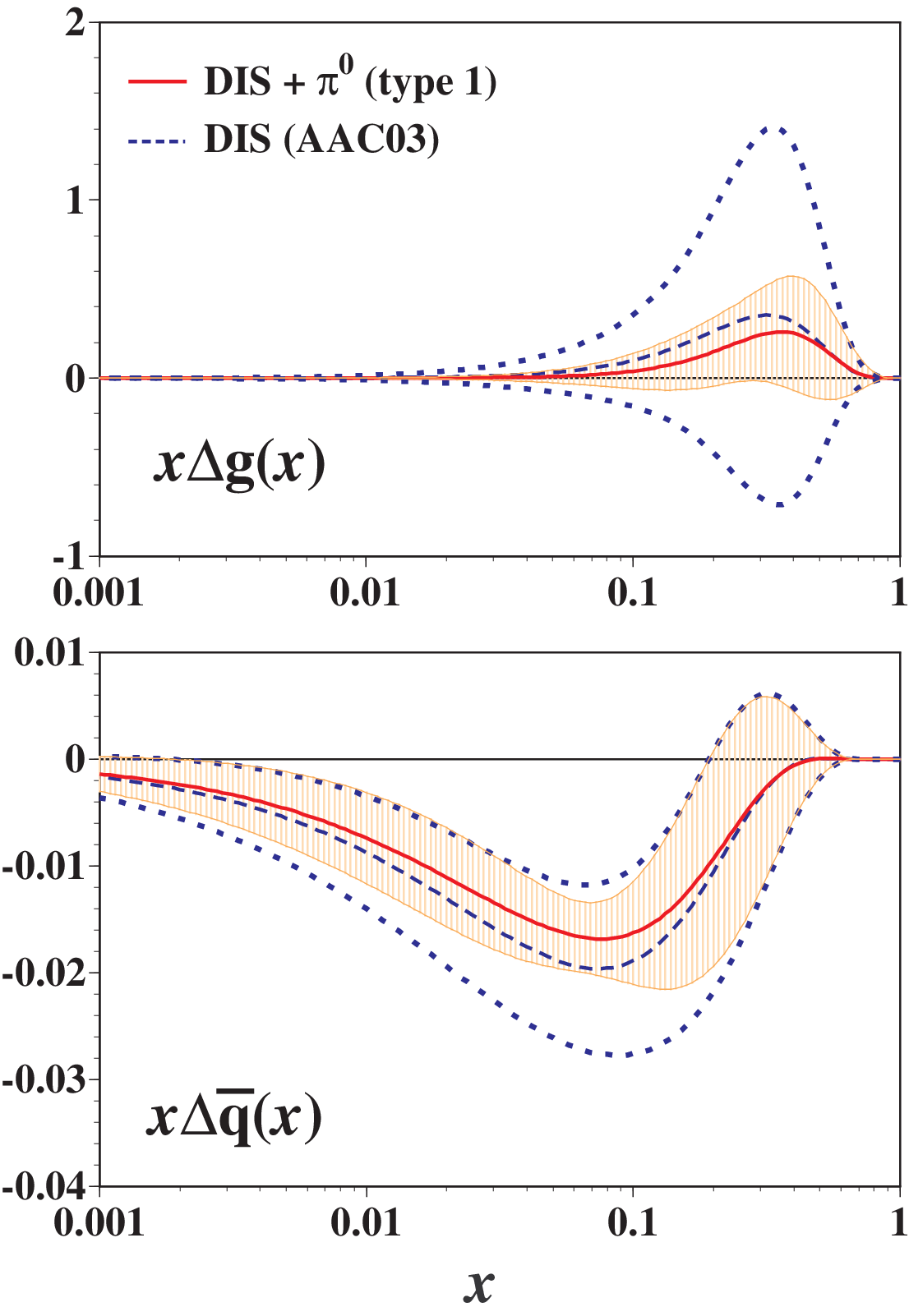}
\caption{\label{fig:exdfvs03} (Color online)
              Polarized PDFs obtained by the type-1 $\chi^2$
              analysis are shown at $Q^2$=1 GeV$^2$ by the solid curves,
              and their uncertainties are shown by the shaded bands.
              In comparison, the AAC03 PDFs are shown by the dashed curves
              and their uncertainties by the dotted curves.
}
\end{figure}

The polarized PDFs obtained by the analyses are shown in
Fig. \ref{fig:exdfvs03} at $Q^2$=1 GeV$^2$, and they are compared
with the AAC03 distributions. Their uncertainties are also shown.
The distributions are almost the same in both analyses;
however, there are much differences between the uncertainty
bands. Although the $\Delta u_v$ uncertainty bands are the same,
the $\Delta d_v$ determination is improved at $x>0.2$ due to
the JLab neutron data. The polarized antiquark distribution is
significantly improved at $x \sim 0.1$ because of {\sc Hermes} and
{\sc Compass} data on the deuteron.

The most significant improvement is found for the polarized gluon
distribution. As clearly shown Fig. \ref{fig:exdfvs03}, the uncertainties
in gluon polarization are significantly reduced in the type-1 fit, where
the $A_{LL}$ for pion production is included. The central curve
also shifted slightly. The AAC03 analysis indicated that a negative
gluon polarization is possible; however, we clearly see some preference
for the positively polarized gluon in the type-1 analysis. We have found
a possibility that $\Delta g(x)$ at large $x$ could be constrained
by the $g_1^d$ data from the {\sc Hermes} and {\sc Compass} collaborations.
The impact of newly obtained data will be further discussed in Sec. \ref{gluon}.

%%%%%%%%%%%%%%%%%%%%%%%%%%%%%%%% figure %%%%%%%%%%%%%%%%%%%%%%%%%%%%%%%%%%%%%%%%
%%% Figure of the uncertainty of the polarized PDF 06vs03 %%%
%%%%%%%%%%%%%%%%%%%%%%%%%%%%%%%% figure %%%%%%%%%%%%%%%%%%%%%%%%%%%%%%%%%%%%%%%%

The first moments of $\Delta \bar{q}(x)$, $\Delta g(x)$, and 
$\Delta \Sigma(x)=\sum_i [q_i(x)+\bar q_i(x)]$
are listed at $Q^2$=1 GeV$^2$ in Table \ref{T:1stm} together with the ones of
the AAC03. Because the quark and antiquark distributions are almost the same
as the AAC03, the first moment of $\Delta \bar q$ and the quark spin content
$\Delta \Sigma$ are similar in the current and AAC03 analyses.
The only difference is the gluon first moment. First, the uncertainty of
$\Delta g$ is significantly reduced in the type-1 analysis
($1.27 \rightarrow 0.32$). This is because of the {\sc Phenix} data,
which constrain the polarized gluon distribution in the region $x \sim 0.1$.
Therefore, the {\sc Phenix} $A_{LL}^{\pi^0}$ measurements play an important role
in the determination of $\Delta g(x)$. However, it should be mentioned
that an effect from the assumed functional form is not included
in estimating the uncertainty.
Furthermore, there is an additional scale uncertainty due to the 
scale error of the {\sc Phenix} data (40\%) originated predominantly 
in the beam polarization uncertainty. This uncertainty is estimated 
to be $\sim$20\% in $\Delta g(x)$ and its first moement, while the 
quark secter remains intact.

%%%%%%%%%%%%%%%%%%%%%%%%%%%%%%%% table  %%%%%%%%%%%%%%%%%%%%%%%%%%%%%%%%%%%%%%%%
\begin{table}[t]
\caption{\label{T:1stm}
The first moments of the obtained polarized PDFs at $Q^2=1$ GeV$^2$. 
The current analysis results are compared with those of the previous
results (AAC03). The $\Delta \Sigma$ is the quark spin content.
}
\begin{ruledtabular}
\begin{tabular}{lccc} 
             & $\Delta \bar{q}$     & $\Delta g$        & $\Delta \Sigma$  \\
\hline
Type 1  & $-$0.05 $\pm$ 0.01 &       0.31 $\pm$ 0.32 & 0.27 $\pm$ 0.07 \\
Type 2  & $-$0.06 $\pm$ 0.02 &       0.47 $\pm$ 1.08 & 0.25 $\pm$ 0.10 \\
AAC03        & $-$0.06 $\pm$ 0.02 &  0.50 $\pm$ 1.27 & 0.21 $\pm$ 0.14
\end{tabular}
\end{ruledtabular}
\end{table}
%%%%%%%%%%%%%%%%%%%%%%%%%%%%%%%% table  %%%%%%%%%%%%%%%%%%%%%%%%%%%%%%%%%%%%%%%%

%%%%%%%%%%%%%%%%%%%%%%%%%%%%%%%% figure %%%%%%%%%%%%%%%%%%%%%%%%%%%%%%%%%%%%%%%%
\begin{figure}[b]
        \includegraphics*[width=40.0mm]{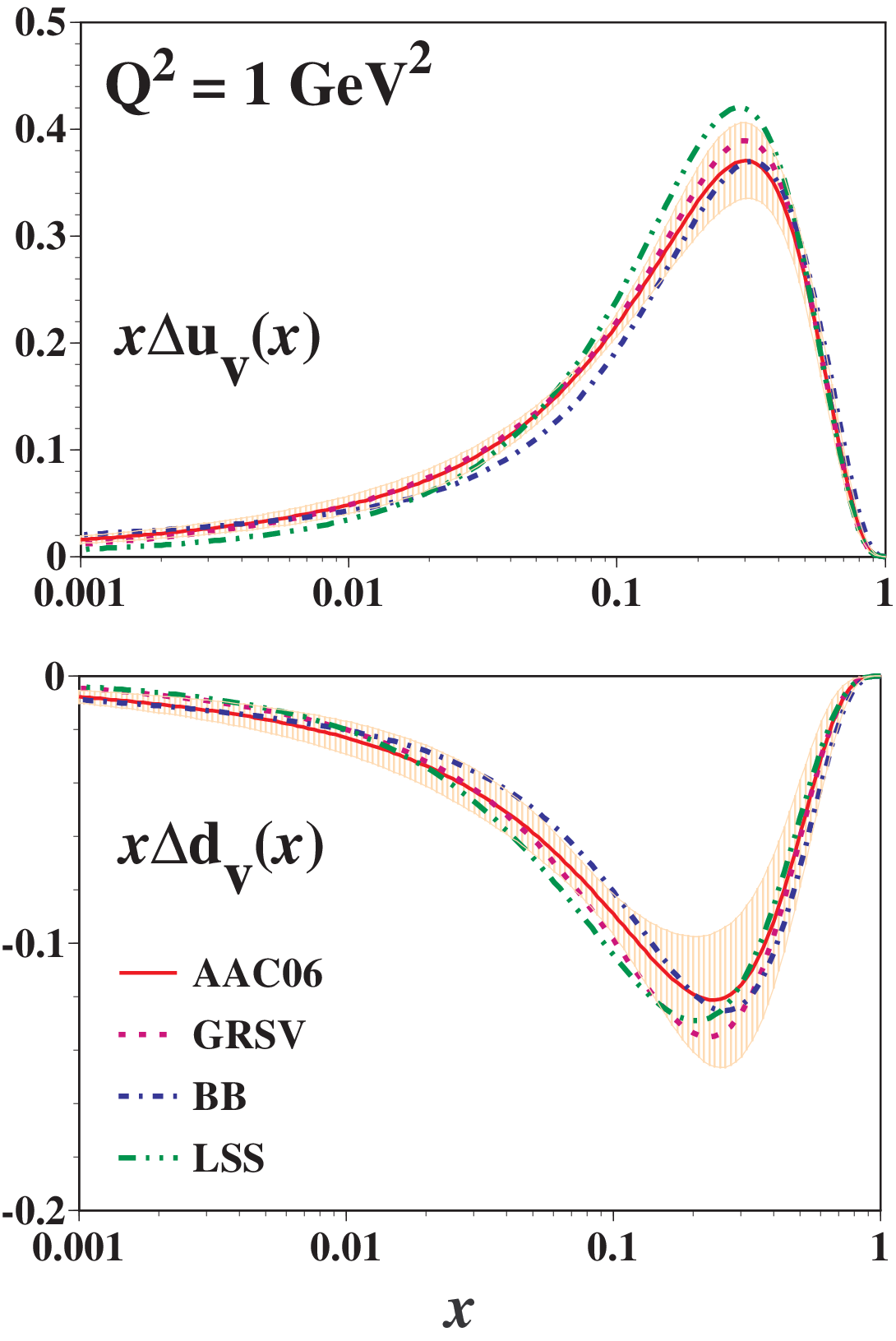} \hspace{1mm}
        \includegraphics*[width=39.5mm]{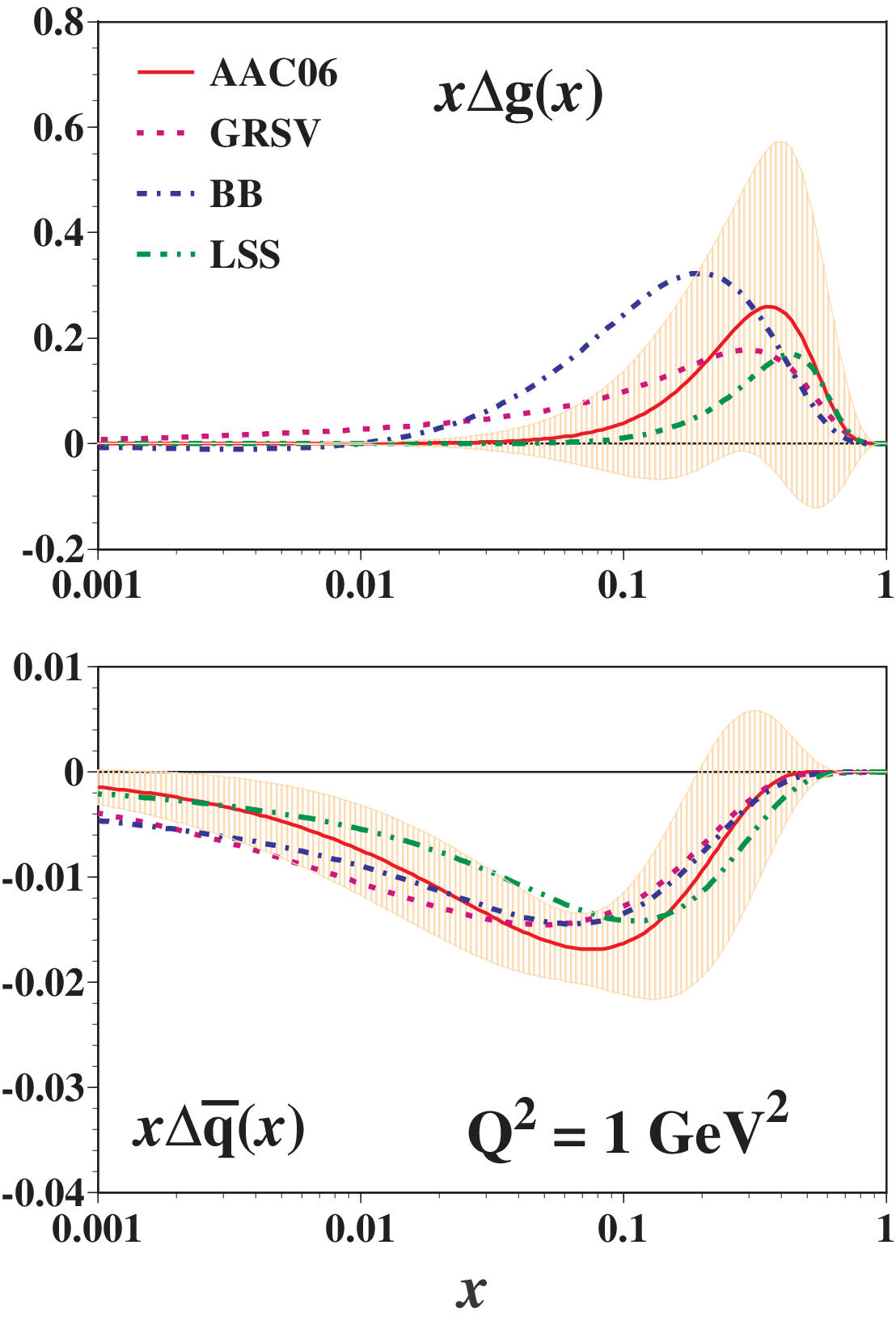} 
\caption{\label{fig:comp} (Color online)
         Comparison with other polarized PDFs at $Q^2$=1 GeV$^2$. 
         The type 1 distributions and their uncertainties are shown by
         the solid curves and bands. The others are the 
         GRSV, BB, and LSS parametrizations.
}
\end{figure}
%%%%%%%%%%%%%%%%%%%%%%%%%%%%%%%% figure %%%%%%%%%%%%%%%%%%%%%%%%%%%%%%%%%%%%%%%%

The polarized PDFs are compared with other recent distributions
at $Q^2$=1 GeV$^2$ in Fig. \ref{fig:comp}. The solid, dashed, dashed-dot,
dotted curves indicate AAC06 (type-1 analysis), 
GRSV (Gl\"uck, Reya, Stratmann, and Vogelsang) \cite{grsv01},
BB (Bl\"umlein and B\"ottcher) \cite{bb02},
and LSS (Leader, Sidorov, and Stamenov) \cite{lss06} distributions,
respectively. All the distributions agree well except for the polarized
gluon distribution. However, the uncertainty band of $\Delta g(x)$
is still huge even if the {\sc Phenix}-pion data are included in the analysis.
The different distributions agree each other because they are roughly within
the uncertainties. The various parametrizations have different analysis
methods such as in the used data set, $x$-dependent functional form,
treatment of higher-twist effects, and positivity condition; however,
the obtained distributions are similar. There are other recent
parametrization studies \cite{recent} with semi-inclusive data by
de Florian, Navarro, and Sassot and also in the statistical-parton-distribution
model by Bourrely, Soffer, and Buccella although they are not shown in 
the figure. There are also other parametrization studies in Ref. \cite{others}.

%%%%%%%%%%%%%%%%%%%%%%%%%%%%%%%% figure %%%%%%%%%%%%%%%%%%%%%%%%%%%%%%%%%%%%%%%%
\begin{figure}[t]
        \includegraphics*[width=60mm]{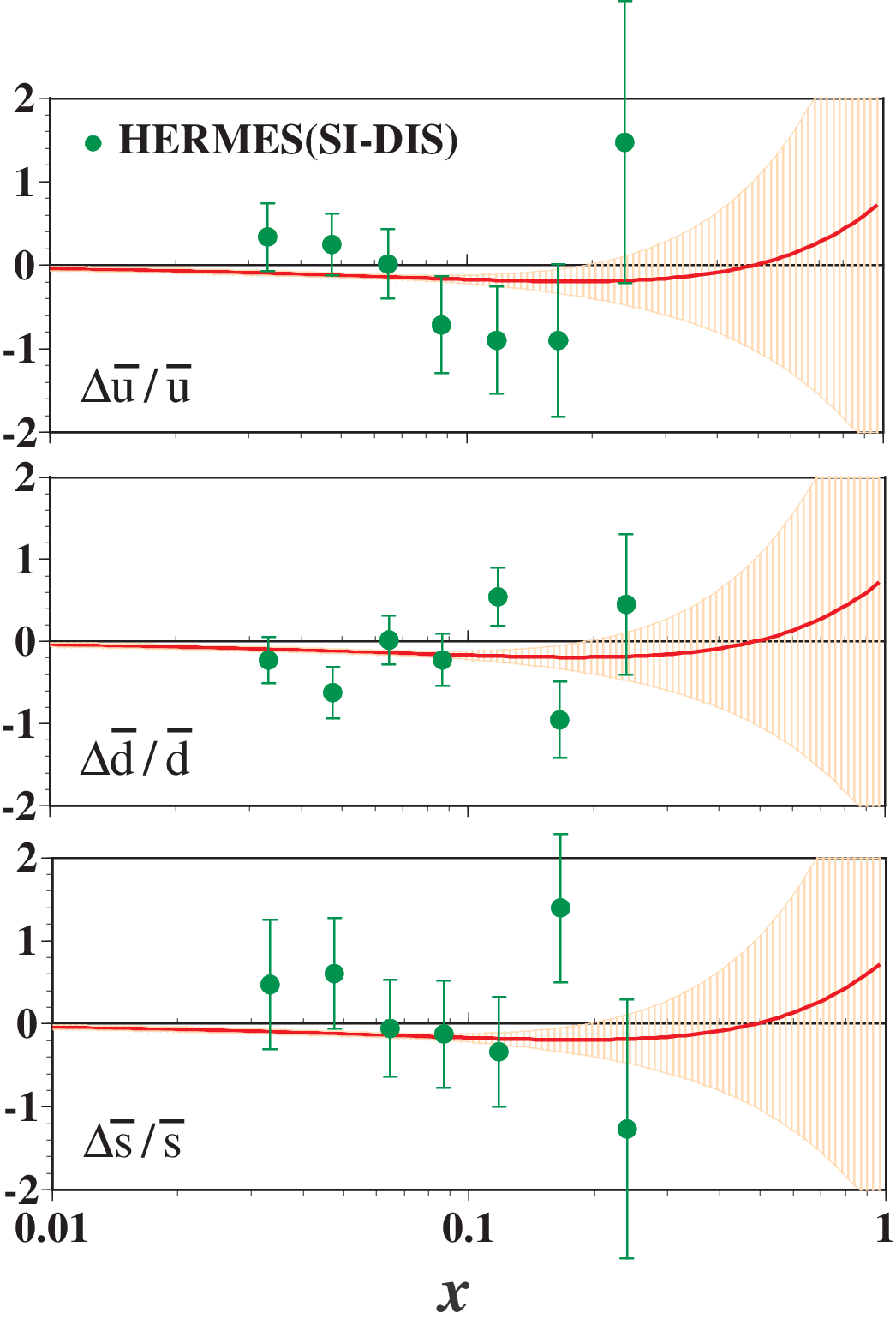}
\caption{\label{fig:sidis-qbar} (Color online)
Comparison with the {\sc Hermes} semi-inclusive DIS data on the antiquark
distribution ratios $\Delta \bar u/\bar u$, $\Delta \bar d/\bar d$,
and $\Delta \bar s/\bar s$.
The solid curves are the type-1 analysis results at $Q^2$=1 GeV$^2$
and the uncertainty ranges are shown by the shaded bands.}
\end{figure}
%%%%%%%%%%%%%%%%%%%%%%%%%%%%%%%% figure %%%%%%%%%%%%%%%%%%%%%%%%%%%%%%%%%%%%%%%%

Next, the polarized antiquark distribution of the type-1 is compared
with experimental antiquark-distribution ratios $\Delta \bar u/\bar u$,
$\Delta \bar d/\bar d$, and $\Delta \bar s/\bar s$ by the {\sc Hermes}
collaboration \cite{hermes-sidis} in Fig. \ref{fig:sidis-qbar}.
The {\sc Hermes} measured semi-inclusive spin asymmetries for pion and kaon
productions in positron and electron DIS with proton and deuteron targets.
They analyzed the semi-inclusive data together with the inclusive ones
on $A_1$ for obtaining the quark and antiquark distribution ratios,
$\Delta q(x)/q(x)$ and $\Delta \bar q (x)/ \bar q(x)$, in the LO.
Effects of NLO radiative corrections are small in comparison with
the experimental errors because the distribution ratios are taken.
Therefore, although the experimental ratios are obtained by LO
analyses and the theoretical ratios are calculated in the NLO,
they can be compared with each other. 
In the AAC analyses, flavor symmetric antiquark distributions are
assumed $\Delta\bar u(x)=\Delta\bar d(x)=\Delta\bar s(x)$ at $Q^2$=1 GeV$^2$; 
however, one should note that the antiquark distributions are different 
($\Delta\bar u(x) \ne \Delta\bar d(x) \ne \Delta\bar s(x)$ at
$Q^2 \ne$1 GeV$^2$) because of $Q^2$ evolution effects in the NLO
\cite{flavor3}. In the figure, the type-1 distributions are shown at
$Q^2$=1 GeV$^2$; however, $Q^2$ variations are small because the ratios are
taken. The theoretical antiquark-distribution ratios agree with the {\sc Hermes}
data if the experimental errors and the parametrization uncertainties are
considered. At this stage, the semi-inclusive data are not included
in the global analysis. The data are fully consistent with our
parametrization. It is obvious that the data would impose significant
constraints when the data precision is improved.

The AAC00 and AAC03 codes are available at the web site \cite{aac03}
for calculating the polarized PDFs. However, such a library is not prepared
for the current analyses because the {\sc Phenix} data utilized here are
still preliminary results. In particular, the data will become more accurate
in the final paper, so that the $\Delta g(x)$ uncertainty should become
smaller. After the data are finalized, we plan to provide a code for the
polarized PDFs.

%%%%%%%%%%%%%%%%%%%%%%%%%%%%%%%%%%%%%%%%%%%%%%%%%%%%%%%%%%%%%%%%%%%%%%%%%%%%%%%%
%%%%%%%%%%%%%%%%%%%%%%%%%%%%%%%%%%%%%%%%%%%%%%%%%%%%%%%%%%%%%%%%%%%%%%%%%%%%%%%%
\section{\label{gluon} Polarized gluon distribution}

In this section, the details are discussed on the determination of
the polarized gluon distribution. First, we explain the possibility
that the recent {\sc Hermes} and {\sc Compass} measurements on $A_1^d$ could
impose a constraint for $\Delta g(x)$ at large $x$ because of
their $Q^2$ differences. Second, the possibility of negative gluon
distribution ($\Delta g(x)<0$) at small $x$ is studied by including
the {\sc Phenix} $\pi^0$ data in the analysis. Third, the polarized gluon
distributions obtained by the current analyses are compared with
{\sc Hermes}, SMC, and {\sc Compass} measurements on the ratio
$\Delta g(x)/g(x)$.

%%%%%%%%%%%%%%%%%%%%%%%%%%%%%%%%%%%%%%%%%%%%%%%%%%%%%%%%%%%%%%%%%%%%%%%%%%%%%%%%
\subsection{\label{large-x-gluon} Polarized gluon distribution at large $x$}

We discuss the possibility that precise measurements of $g_1$ at $x \sim 0.05$
could impose a constraint on $\Delta g(x)$ at larger $x$. As shown
in Eq. (\ref{eqn:g1}), the structure function is given by the convolution
integral of the polarized PDFs with corresponding coefficient functions. 
In particular, we discuss an effect of the gluon term
$(x/z) \Delta g(x/z) \Delta C_g(z)$, where $z$ is the integration variable 
in Eq. (\ref{eqn:conv}).

The parametrization results of the type 1 are compared with inclusive DIS
data particularly by JLab, {\sc Hermes}, and {\sc Compass} on the spin
asymmetry $A_1$. Then, the parametrization is consistent with the SLAC-E154
neutron data and {\sc Compass} deuteron data. Although the agreement is
excellent also with the {\sc Hermes} proton data for $A_1^p$, the curve deviates
at small $x$ ($0.03<x<0.07$) in the {\sc Hermes} deuteron data ($A_1^d$).

%%%%%%%%%%%%%%%%%%%%%%%%%%%%%%%% figure %%%%%%%%%%%%%%%%%%%%%%%%%%%%%%%%%%%%%%%%
\begin{figure}[b]
        \includegraphics*[width=82mm]{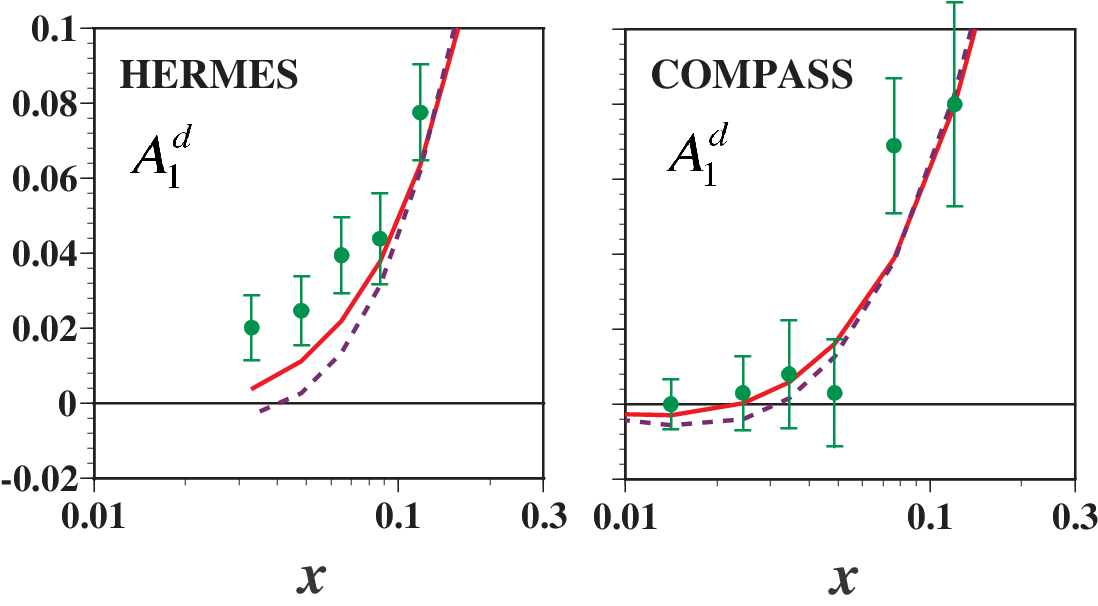}
\caption{\label{fig:A1dHvsC} (Color online)
     The {\sc Hermes} and {\sc Compass} data on the spin asymmetry of
     the deuteron ($A_1^d$) at small $x$ are shown in comparison with
     the parametrization results. The solid curves indicate the type-1
     asymmetries and the dashed ones are obtained simply by removing
     the NLO gluon term ($\Delta C_g=0$) in the structure function $g_1$.
}
\end{figure}
%%%%%%%%%%%%%%%%%%%%%%%%%%%%%%%% figure %%%%%%%%%%%%%%%%%%%%%%%%%%%%%%%%%%%%%%%%

It is noteworthy that the central curve of our fit results deviates from
{\sc Hermes} $A_1^{d}$ at small $x$, although a reasonable agreement
is obtained with {\sc Compass} $A_1^{d}$ in the same $x$-region.
The deviation from the {\sc Hermes} data points becomes larger when
we artificially remove the gluon contribution to $g_1^{d}$, which is
shown as dashed curves in Fig. \ref{fig:A1dHvsC}.
Because the $Q^2$ values are quite different: $Q^2_{\text{\rm{\sc Hermes}}}=
1$~GeV$^2$ and $Q^2_{\text{\rm{\sc Compass}}}= 6$~GeV$^2$, we could
attribute the observed discrepancy to the $Q^2$ difference.
There are two possible effects of such kind; one is the higher-twist
effect as pointed out recently in Ref. \cite{lss06}. The other, which we
found in this analysis, is due to {\it a positive
gluon polarization at large $x$}, which we discuss below.

\begin{figure}[t]
        \includegraphics*[width=60mm]{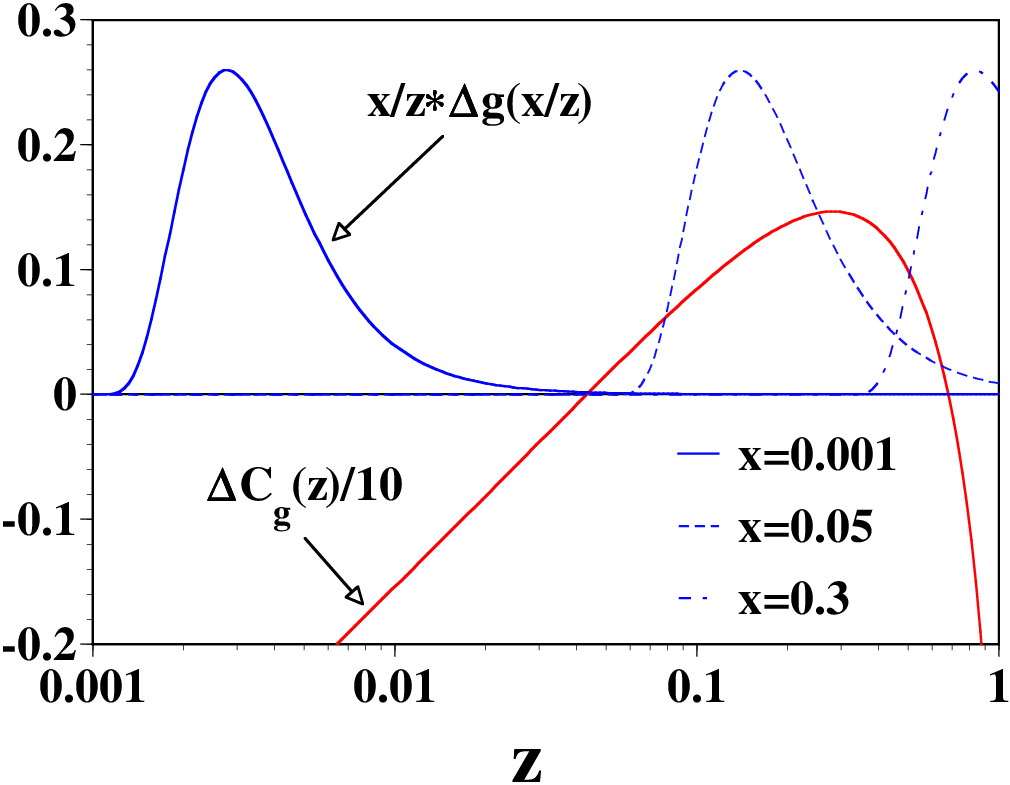}
\caption{\label{fig:corr1} (Color online)
       The polarized gluon distributions and coefficient function
       are shown at $x$=0.001, 0.05, and 0.3 for the type-1 parametrization.
}
\end{figure}

We recall the gluon contribution to the $g_1$ is 
$(x/z) \Delta g(x/z) \Delta C_g(z)$ integrated with respect to
$z$ over $(x,1)$. In this picture, the gluon with the momentum
fraction $x/z$ would split into the quark with a momentum fraction
$x$ and anything else to contribute to the $g_1(x)$.
Figure \ref{fig:corr1} displays the type-1 polarized gluon distribution
$(x/z) \Delta g(x/z)$ for selected $x$-values and
the polarized coefficient function $\Delta C_g(z)$ as functions of $z$.
The $\Delta C_g(z)$ is positive in the region, $0.05 < z<0.7$,
whereas it is negative in other $z$ region.
The function $(x/z) \Delta g(x/z)$ for $x=0.001$ peaks at around
$z=0.002$ and it moves to a larger $z$ as $x$ becomes larger
($x$=0.05 and 0.3). Therefore, the gluon contribution to the $g_1$
is negative in the small-$x$ region, positive at medium $x$ ($\sim 0.05$),
and back to negative at large $x$.

%%%%%%%%%%%%%%%%%%%%%%%%%%%%%%%% figure %%%%%%%%%%%%%%%%%%%%%%%%%%%%%%%%%%%%%%%%
%%%%%%%%%%%%%%%%%%%%%%%%%%%%%%%% figure %%%%%%%%%%%%%%%%%%%%%%%%%%%%%%%%%%%%%%%%

The {\sc Hermes} data shows a discrepancy from our
parametrization result with the gluon contribution eliminated
especially in the region, 0.033$\le x \le$0.065. This is the
$x$ region where the gluonic contribution from $\Delta g(x) > 0$
is positive. If the $\Delta g$ is negative, the gluonic
contribution to $g_1$ will be negative so that the discrepancy
becomes larger. Therefore, we could interpret the deviation
of {\sc Hermes} data as an indication of positive $\Delta g(x)$
at large $x$. 

Obviously further precision studies are necessary to identify
the source of the discrepancy. Determination of the sign of the gluon
polarization will be helpful. More precise data on the $Q^2$ dependence
of $g_1$ would be certainly helpful, too.

%%%%%%%%%%%%%%%%%%%%%%%%%%%%%%%%%%%%%%%%%%%%%%%%%%%%%%%%%%%%%%%%%%%%%%%%%%%%%%%%
\subsection{\label{small-x-gluon} Polarized gluon distribution at small $x$}

%%%%%%%%%%%%%%%%%%%%%%%%%%%%%%%% figure %%%%%%%%%%%%%%%%%%%%%%%%%%%%%%%%%%%%%%%%
\begin{figure}[b]
        \includegraphics*[width=60mm]{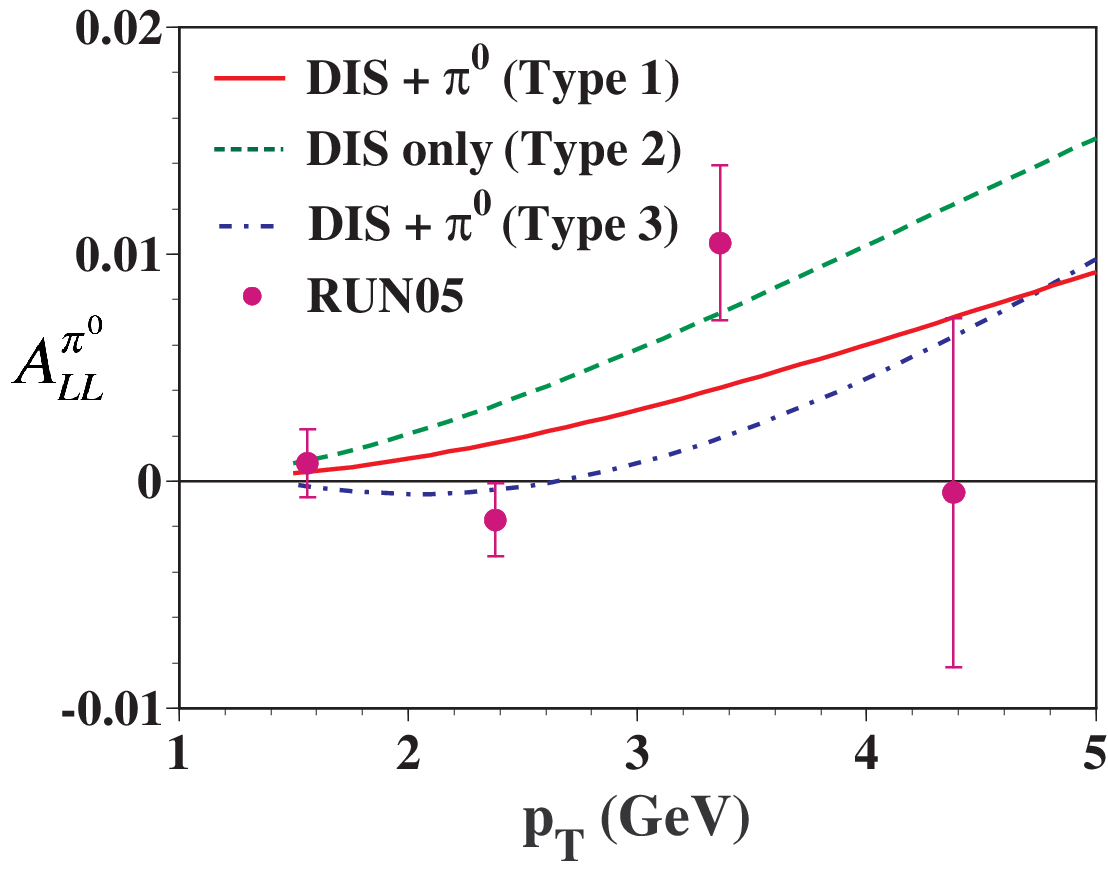}
\caption{\label{fig:negative-dg} (Color online)
Theoretical asymmetries are shown for the type 1, 2, and 3 parametrizations
in comparison with the {\sc Phenix} data of $A_{LL}^{\pi^0}$. Only
the small-$p_T$ region is shown in order to see the differences between
the theoretical asymmetries at small $p_T$.
}
%\end{figure}
%%%%%%%%%%%%%%%%%%%%%%%%%%%%%%%% figure %%%%%%%%%%%%%%%%%%%%%%%%%%%%%%%%%%%%%%%%
%
\vspace{0.7cm}
%%%%%%%%%%%%%%%%%%%%%%%%%%%%%%%% figure %%%%%%%%%%%%%%%%%%%%%%%%%%%%%%%%%%%%%%%%
%\begin{figure}[b]
        \includegraphics*[width=60mm]{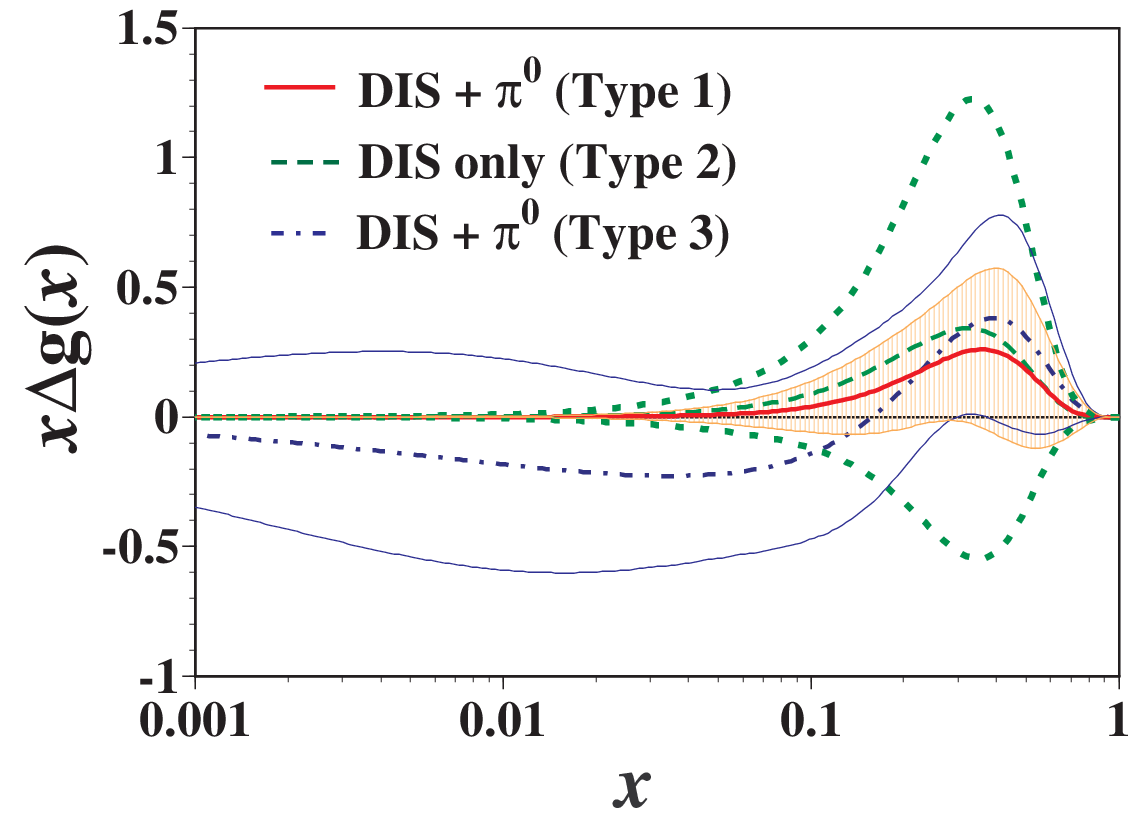}
\caption{\label{fig:wg2} (Color online)
    The polarized gluon distributions are shown at $Q^2$=1 GeV$^2$. 
    The solid, dashed, and dotted curves indicate the type 1, 2, and 3
    distributions, respectively. 
    The uncertainty ranges are shown by the shaded band, the dotted curves,
    and the thin solid curves for the type 1, 2 and 3, respectively.
}
\end{figure}
%%%%%%%%%%%%%%%%%%%%%%%%%%%%%%%% figure %%%%%%%%%%%%%%%%%%%%%%%%%%%%%%%%%%%%%%%%

We have discussed the polarized gluon distribution at large $x$
basing on $Q^2$ dependence of $A_1^d$. On the other hand,
the gluon polarization in small-$x$ region is not obvious.
Most of the global analyses showed $\Delta g(x) > 0$ in the small-$x$
region, and estimated uncertainty turned out to be small including our
previous analysis, BB, and LSS.@We are going to show that different
small-$x$ behavior is still allowed within the current framework,
and that the uncertainty band changes drastically.

We start from the possibility of negative gluon polarization
inspired from the two possible solutions of $\Delta g(x)/g(x)$
to reproduce $A_{LL}^{\pi^0}$ measured by {\sc Phenix}.
Since the pion production is dominated by $gg$ scattering at small $p_T$,
the asymmetry $A_{LL}^{\pi^0}$ approximately depends
on quadratic function of $\Delta g(x)/g(x)$. Indeed, two solutions,
positive and negative $\Delta g(x)/g(x)$ can be obtained, when
only $A_{LL}^{\pi^0}$ data are used to constrain
$\Delta g(x)/g(x)$ \cite{a1-pi-theo,Saito-PANIC}. The {\sc Phenix} 
$A_{LL}^{\pi^0}$ is sensitive to the gluon around $x \sim 0.1$,
and the negative $\Delta g(x)/g(x)$ solution can coexists
with the previously mentioned positive $\Delta g(x)$ at large $x$.

To explore this possibility, we have performed another global fit
by assigning the initial distribution, $\Delta g(x)=-\Delta g(x)_{\rm type-1}$,
at $Q_0^2$ without modifying the parametrization in Eq. (\ref{eqn:df}).
The fit was successful, and there was no practical difference 
in $\chi^2$. The resulted parameter set is referred to as type 3.
The fit results are shown in Fig. \ref{fig:negative-dg} in comparison
with the type-1 and 2 parametrizations and the {\sc Phenix} experimental
data on the double spin asymmetry $A_{LL}^{\pi^0}$. 
For explaining the negative asymmetry at $p_T$=2.38 GeV, a significant
change of the functional form of $\Delta g(x)$ is needed in the region,
$0.06<x<0.2$. There is no noticeable change in the quark
distributions and only $x \Delta g(x)$ is shown in Fig. \ref{fig:wg2}.
As can be expected from the discussion above, the $\Delta g(x)$
is negative up to $x \sim 0.1$ and retains positive value towards
larger $x$ region especially to reproduce $Q^2$ dependence of
$A_1^{d}$.

Three types of the gluon polarization $\Delta g(x)$ are displayed with
their uncertainties in Fig. \ref{fig:wg2}.
The uncertainty bands are reduced in the analyses with the pion data.
In principle, $\Delta g(x)$ should be constrained only in the PHENIX
kinematical range $x \sim 0.1$. However, the reduction is also found
in the region $x>0.2$. This is because a smooth functional form is
assumed in the analysis, where the constraint in the $x \sim 0.1$
region also affects the uncertainty estimation in other $x$ regions.
Because the gluon-gluon subprocesses dominate, the polarized cross
section for pion-production depends on $\Delta g$ quadratically.
It leads to two independent solutions, the type 1 and type 3. 
In the region $x \gtrsim 0.1$ partially covered by the PHENIX data,
similar uncertainties are obtained and they cover both solutions.
However, the type-3 $\Delta g(x)$ resides outside of the uncertainty
bands of type 1 and type 2 at small $x$, which reveals that the uncertainty
depends on the functional behavior of $\Delta g(x)$. The errors due to
this functional form are not included in showing the uncertainty bands
in our analysis and also other ones \cite{bb02,lss06}, because there is
no systematic way to include them.

The large uncertainty in type-3 $\Delta g(x)$
in the smaller-$x$ region is also reflected to the first moment:
$-0.56 \pm 2.16$. The large error comes from the small-$x$ region,
where there is no experimental data which constrains the small-$x$
behavior of $\Delta g(x)$. In fact, if the type-3 $\Delta g(x)$
is integrated over the region, $0.1 \le x \le 1$, covered by
the DIS and pion-production data, the first moment becomes
$0.32 \pm 0.42$ which is comparable to the type-1 value,
$0.30 \pm 0.30$, in the same $x$ region.
Only the $Q^2$ evolution and smoothness of functional forms are
the possibilities to provide some constraints in the small-$x$ region. 
Please note that there is no direct experimental constraint for
$\Delta g(x)$ at small $x$, therefore it is highly unconstrained.
In order to clarify the situation, we need a definitive
measurement of the sign of the gluon polarization around
$x \lesssim 0.1$. The higher energy run ($\sqrt{s} = 500$~GeV)
at RHIC and eRHIC/e-LIC would play important roles in such measurements.

%%%%%%%%%%%%%%%%%%%%%%%%%%%%%%%%%%%%%%%%%%%%%%%%%%%%%%%%%%%%%%%%%%%%%%%%%%%%%%%%
\subsection{\label{comp-high-pt} 
             Comparison with high-$p_T$ hadron-production data in DIS}

There are reports on the polarized gluon distribution from high-$p_T$
hadron production in DIS by {\sc Hermes} \cite{hermes-dg}, 
SMC \cite{smc-dg}, and {\sc Compass} \cite{compass-dg-lowq2} collaborations.
Their data are shown by the ratio $\Delta g(x)/g(x)$.
There are also preliminary data by the {\sc Compass} \cite{compass-dg-preli}
for $\Delta g(x)/g(x)$ from analyses of high-$Q^2$ ($>$1 GeV$^2$)
hadron-production and charmed-hadron-production data. 

\begin{figure}[h]
        \includegraphics*[width=60mm]{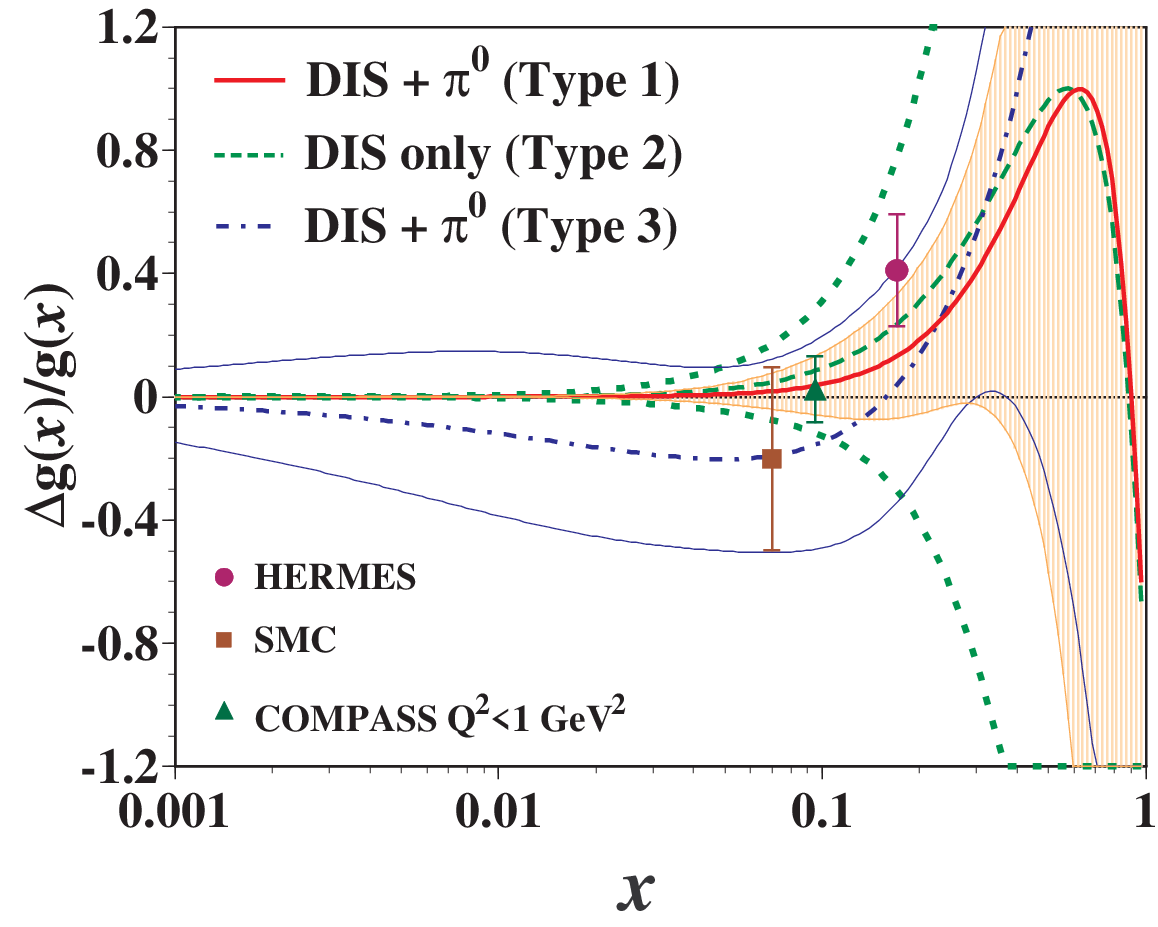}
\caption{\label{fig:wg} (Color online)
     The theoretical ratios $\Delta g(x)/g(x)$ of the type 1, 2, and 3
     are compared with high-$p_T$ hadron-production data measured
     in DIS by {\sc Hermes}, SMC, and {\sc Compass} collaborations.
     The solid, dashed, and dashed-dot curves indicate type 1, 2, and 3
     parametrization results, respectively.
     The uncertainty ranges are shown by the shaded band, the dotted curves,
     and the thin solid curves for the type 1, 2 and 3, respectively.
}
\end{figure}

We compare our results with those experimental data in Fig. \ref{fig:wg}.
The parametrization results for the type 1, 2, and 3 are shown together
with their uncertainty bands. The high-$p_T$ hadron
data in DIS impose a constraint for the polarized gluon distribution
in the $x$ range, $0.07<x<0.2$. We find that three fit results are 
consistent with these data within both uncertainties.

It is interesting to note that the $\Delta g(x)/g(x)$
distributions obtained by the analyses are similar to a theoretical
prediction by Brodsky and Schmidt \cite{bs90} in the region, $x\sim$0.1.

%%%%%%%%%%%%%%%%%%%%%%%%%%%%%%%% figure %%%%%%%%%%%%%%%%%%%%%%%%%%%%%%%%%%%%%%%%
%%%%%%%%%%%%%%%%%%%%%%%%%%%%%%%% figure %%%%%%%%%%%%%%%%%%%%%%%%%%%%%%%%%%%%%%%%

%%%%%%%%%%%%%%%%%%%%%%%%%%%%%%%%%%%%%%%%%%%%%%%%%%%%%%%%%%%%%%%%%%%%%%%%%%%%%%%%
%%%%%%%%%%%%%%%%%%%%%%%%%%%%%%%%%%%%%%%%%%%%%%%%%%%%%%%%%%%%%%%%%%%%%%%%%%%%%%%%
\section{\label{summary} Summary}

We have performed the global analysis of spin asymmetry data from DIS and
neutral-pion production in polarized $pp$ scattering. First, we
showed two types of analysis results by using DIS and {\sc Phenix} pion data
and by using only the DIS data. In comparison with the previous version
(AAC03), new data are added from the JLab $A_1^n$, {\sc Hermes} and
{\sc Compass} $A_1^d$, and {\sc Phenix} $A_{LL}^{\pi^0}$ measurements.
We found that the JLab neutron data contribute to reduce the uncertainty
band of the valence distribution $\Delta d_v$, the {\sc Hermes} and
{\sc Compass} data to improve the antiquark distribution $\Delta \bar q$
in the region, $x\sim 0.1$. 

There are two important consequences on the polarized gluon distribution.
First, the differences between the {\sc Hermes} and {\sc Compass} $A_1^d$
data could indicate a positive gluon polarization at large $x$ because of
their $Q^2$ differences. Second, the {\sc Phenix} $A_{LL}^{\pi^0}$
measurements improve the determination of $\Delta g(x)$ significantly.
In fact, the uncertainty band of $\Delta g(x)$ becomes smaller by 60\%.

\vspace{0.5cm}
%%%%%%%%%%%%%%%%%%%%%%%%%%%%%%%%%%%%%%%%%%%%%%%%%%%%%%%%%%%%%%%%%%%%%%%%%%%%%%%%
%%%%%%%%%%%%%%%%%%%%%%%%%%%%%%%%%%%%%%%%%%%%%%%%%%%%%%%%%%%%%%%%%%%%%%%%%%%%%%%%
\begin{acknowledgements}
This work is supported by the Grant-in-Aid for Scientific Research from
the Japanese Ministry of Education, Culture, Sports, Science, and Technology.
It is also supported by RIKEN and the Japan-U.S. Cooperative Science Program. 
The authors thank K. Sudoh and the entire RHIC Spin Collaboration,
especially A. Bazilevsky,  A. Deshpande, Y. Fukao, M. Stratmann,
and W. Vogelsang, for helpful discussions on the gluon polarization.
\end{acknowledgements}

%%%%%%%%%%%%%%%%%%%%%%%%%%%%%%%%%%%%%%%%%%%%%%%%%%%%%%%%%%%%%%%%%%%%%%%%%%%%%%%%
%%%%%%%%%%%%%%%%%%%%%%%%%%%%%%%%%%%%%%%%%%%%%%%%%%%%%%%%%%%%%%%%%%%%%%%%%%%%%%%%
% \appendix
% \section{\label{library} Practical code for calculating polarized
%          parton distribution functions}

%%%%%%%%%%%%%%%%%%%%%%%%%%%%%% reference %%%%%%%%%%%%%%%%%%%%%%%%%%%%%%%%%%%%%%%%%%%%

%%%%%%%%%%%%%%%%%%%%%%%%%%%%%% reference %%%%%%%%%%%%%%%%%%%%%%%%%%%%%%%%%%%%%%%%%%%%


\begin{thebibliography}{00}
\newcommand{\etal}{{\it et al.}}
\bibitem{aac00} Asymmetry Analysis Collaboration (AAC),
                Y. Goto, N. Hayashi, M. Hirai, H. Horikawa, S. Kumano,
                M. Miyama, T. Morii, N. Saito, T.-A. Shibata, E. Taniguchi,
                T. Yamanishi,
                     Phys. Rev. {\bf D62}, 034017 (2000).
\bibitem{aac03} Asymmetry Analysis Collaboration (AAC),
                M. Hirai, S. Kumano, N. Saito (AAC),
                     Phys. Rev. {\bf D69}, 054021 (2004).
                The AAC code for polarized PDFs is available at
                     http://spin.riken.bnl.gov/aac/.
\bibitem{hermes-dg} HERMES collaboration, A. Airapetian {\it et al.},
                                 Phys. Rev. Lett. {\bf 84}, 2584 (2000).
\bibitem{smc-dg}    Spin Muon Collaboration (SMC), B. Adeva {\it et al.},
                                 Phys. Rev. {\bf D70}, 012002 (2004).
\bibitem{compass-dg-lowq2}  {\sc Compass} collaboration, E. S. Ageev {\it et al.},
                                  hep-ex/0511028.
\bibitem{compass-dg-preli}  F. Bradamante, talk given at the workshop
                            on Hadron Structure at J-PARC, Nov. 30 - Dec. 2, 2005,
                            KEK, Tsukuba, Japan,  
                            http://www-conf.kek.jp/J-PARC-HS05/.
\bibitem{phenix-pi-04} {\sc Phenix} collaboration, S. S. Adler {\it et al.},
                       Phys. Rev. Lett. {\bf 93}, 202002 (2004).
\bibitem{phenix-pi-06} {\sc Phenix} collaboration, 
                       K. Boyle,  talk at the XVIIth Particles
                       and Nuclei International Conference (PANIC05),
                       http://www.panic05.lanl.gov/ .
\bibitem{star-jet} {\sc Star} collaboration, J. Kiryluk, hep-ex/0512040.
\bibitem{jlab-g1n}    Jefferson Lab Hall A collaboration, X. Zheng \etal,
                         Phys. Rev. Lett {\bf 92}, 012004 (2004)
\bibitem{hermes-g1d} {\sc Hermes} collaboration,
                         A. Airapetian \etal, Phys. Rev. {\bf D71}, 012003 (2005).
\bibitem{compass-g1d} {\sc Compass} collaboration, E. S. Ageev \etal,
                         Phys. Lett. {\bf B612}, 154 (2005).
\bibitem{a1} R. Devenish and A. Cooper-Sarkar, {\it Deep Inelastic Scattering},
                         Oxford University press (2004). See also, for example,
                         Ref. \cite{emc}.
\bibitem{g1} Expressions of NLO corrections are found, for example, in 
             M. Hirai, S. Kumano, and M. Miyama,
                         Comput. Phys. Commun. {\bf 108}, 38 (1998).
\bibitem{a1-pi-theo} B. J\"ager, A. Sch\"after, M. Stratmann, and W. Vogelsang,
                         Phys. Rev. {\bf D67}, 054005 (2003);
                     B. J\"ager, M. Stratmann, S. Kretzer, and W. Vogelsang,
                         Phys. Rev. Lett. {\bf 92}, 121803 (2004);
                     M. Hirai and K. Sudoh, Phys. Rev. {\bf D71}, 014022 (2005).              
\bibitem{subprocess} E. Leader, {\it Spin in Particle Physics},
                            Cambridge University Press (2001).
\bibitem{a1-pi-kine} R. D. Field, {\it Applications of perturbative QCD},
                     Addison-Wesley Publishing Company (1989).
\bibitem{CTEQ-K}     J. Pumplin, D. R. Stump, J. Huston, H. L. Lai, P. Nadolsky, 
                     and W. K. Tung, JHEP, {\bf 0207}, 012 (2002);
                     See T. Kluge, K. Rabbertz, and M. Wobisch,
                     talk at the TeV4LHC workshop, CERN, Switzerland, 
                     April 28-30, 2005.
\bibitem{afr98} G. Altarelli, S. Forte, and G. Ridolfi,
                     Nucl. Phys. {\bf B534}, 277 (1998).
\bibitem{otherpdfs} A. D. Martin, R. G. Roberts, W. J. Stirling, and R.S. Thorne,
                       Phys. Lett. {\bf B604}, 61 (2004);
                    S. Alekhin, JETP Lett. {\bf 82}, 628 (2005).
\bibitem{dglap-evol} $Q^2$ evolution programs are discussed in Ref. \cite{g1}
                     and M. Miyama and S. Kumano,
                         Comput. Phys. Commun. {\bf 94}, 185 (1996).             
\bibitem{minuit} F. James, CERN Program Library Long Writeup D506. See
              http://wwwasdoc.web.cern.ch/wwwasdoc/minuit \\ /minmain.html.
\bibitem{hessian} J. Pumplin \etal,
                       Phys. Rev. {\bf D65}, 014013 (2002).
\bibitem{del-chi2} For example, see 
                    http://wwwasdoc.web.cern.ch/wwwasdoc \\ /minuit/node33.html,
                    http://www.library.cornell.edu/nr \\ /bookfpdf/f15-6.pdf,
                    http://ccwww.kek.jp/pdg/2005/ \\ reviews/statrpp.pdf.
                    A different  choice of $\Delta \chi^2$ is discussed in 
                    J. C. Collins and J. Pumplin, hep-ph/0105207.
\bibitem{SLAC-R} L. W. Whitlow, S. Rock, A. Bodek, S. Dasu, and E. M. Riordan,
                    Phys. Lett. {\bf B250}, 193 (1990);
                L. W. Whitlow, report SLAC-0357 (1990);
                K. Abe \etal, Phys. Lett. {\bf B452}, 194 (1999).
\bibitem{GRV98}  M. Gl\"uck, E. Reya, and A. Vogt,
                   Eur. Phys. J. {\bf C5}, 461 (1998).
\bibitem{emc} European Muon Collaboration (EMC), J. Ashman \etal,
              Phys. Lett. {\bf B206}, 364 (1988);
              Nucl. Phys. {\bf B328}, 1 (1989).
\bibitem{smc} Spin Muon Collaboration (SMC), B. Adeva \etal, 
               Phys. Rev. {\bf D58}, 112001 (1998).
\bibitem{slac} SLAC-E130 collaboration, G. Baum \etal,
                 Phys. Rev. Lett. {\bf 51}, 1135 (1983);
               E142 collaboration, P. L. Anthony \etal, 
                 Phys. Rev. {\bf D54}, 6620 (1996);
               E143 collaboration, K. Abe \etal,   
                 Phys. Rev. {\bf D58}, 112003 (1998);
               E154 collaboration, K. Abe \etal, 
                 Phys. Rev. Lett. {\bf 79}, 26 (1997);
               E155 collaboration, P. L. Anthony \etal, 
                 Phys. Lett. {\bf B463}, 339 (1999);
                 Phys. Lett. {\bf B493}, 19 (2000).
\bibitem{hermes97} {\sc Hermes} collaboration,
               K. Ackerstaff \etal, Phys. Lett. {\bf B404}, 383 (1997).
\bibitem{kkp} B. A. Kniehl, G. Kramer, and B. P\"otter,
                Nucl. Phys. {\bf B582}, 514 (2000).
\bibitem{RBRC-WS} W. Vogelsang, in BNL-52635,
             Proceedings of RIKEN BNL Research Center Workshop Vol 33,
             Spin Physics at RHIC in Year-1 and Beyond.
\bibitem{hera} ZEUS Collaboration, S. Chekanov \etal,
                  Eur. Phys. J. {\bf C21}, 443 (2001).
\bibitem{elic}  See http://casa.jlab.org/research/elic/elic.shtml .
\bibitem{erhic} See http://www.phenix.bnl.gov/WWW/publish/abhay \\
                 /Home\_of\_EIC/ .
\bibitem{grsv01}  M. Gl\"uck, E. Reya, M. Stratmann, and W. Vogelsang,
                       Phys. Rev. {\bf D63}, 094005 (2001).
\bibitem{bb02}    J. Bl\"umlein and H. B\"ottcher,
                       Nucl. Phys. {\bf B636}, 225 (2002).
\bibitem{lss06}  E. Leader, A. V. Sidorov, and D. B. Stamenov,
                       Phys. Rev.   {\bf D73}, 034023 (2006).
\bibitem{recent} D. de Florian, G. A. Navarro, and R. Sassot,
                       Phys. Rev. {\bf D71},  094018 (2005);
                 C. Bourrely, J. Soffer, and F. Buccella,
                       Eur. Phys. J. {\bf C41}, 327 (2005).
\bibitem{others} T. Gehrmann and W. J. Stirling,
                  Phys. Rev. {\bf D53}, 6100 (1996);
                 G. Altarelli, R. D. Ball, S. Forte, and G. Ridolfi,
                  Nucl. Phys. {\bf B496}, 337 (1997);
                  Acta Phys. Pol. {\bf B29}, 1145 (1998);
                 L. E. Gordon, M. Goshtasbpour, and G. P. Ramsey, 
                  Phys. Rev. {\bf D58}, 094017 (1998).       
\bibitem{hermes-sidis} {\sc Hermes} collaboration, A. Airapetian \etal, 
                        Phys. Rev {\bf D71}, 012003 (2005).
\bibitem{flavor3} S. Kumano, Phys. Rep. {\bf 303}, 183 (1998);
                  G. T. Garvey and J.-C. Peng,
                       Prog. Part. Nucl. Phys. {\bf 47}, 203 (2001).
\bibitem{Saito-PANIC} N. Saito, talk at the
                  Particles and Nuclei International Conference
                  Santa Fe, U.S.A, October 24-28, 2005. 
\bibitem{bs90} S. J. Brodsky and I. Schmidt, 
                       Phys. Lett. {\bf 234}, 144 (1990). 
\end{thebibliography}
\end{document}